\documentclass[12pt]{article}
\usepackage{amsmath,verbatim,color,amssymb,epsfig,amsfonts,natbib}
\numberwithin{equation}{section}
\begin{document}
\newtheorem{lemma}{Lemma}
\newtheorem{theorem}{Theorem}
\newtheorem{remark}{Remark}
\newtheorem{assumption}{Assumption}

\begin{center}{\large \bf
Statistical inference of
heterogeneous treatment effects using semiparametric single-index model
}
\end{center}

\begin{center}
{
{Jichang Yu$^{1}$, Wenjing Chang$^{1}$, Peichao Yu$^{1}$, Lijun Chen$^{2}$, Yuanshan Wu$^{1,\dag}$\\
$^{1}$School of Statistics and Mathematics, Zhongnan University of Economics and Law,\\
Wuhan, Hubei 430073,  China\\
$^{2}$Center for Applied Mathematics of Guangxi,
Guangxi Colleges and Universities Key Laboratory of Complex System  Optimization and Big Data Processing,
Yulin Normal  University,  Yulin 537000, China\\

{\it Email:} wu@zuel.edu.cn
}}
\end{center}

\noindent {\bf Abstract}\quad
In recent years, with the rapid development of science and technology, heterogeneous treatment effects  have emerged as a focal research topic in statistics, econometrics, and sociology. This paper investigates HTE through semiparametric single-index models based on doubly robust estimation. Departing from conventional approaches, we neither impose boundedness constraints on the link function in single-index models nor restrict its support range. By employing the sieve method to approximate the link function, we achieve simultaneous estimation of both the link function and index parameters. Our study not only establishes the asymptotic properties of the proposed estimator but also systematically evaluates its finite-sample performance through comprehensive simulation studies. Numerical results demonstrate that our method significantly outperforms other commonly used competing estimators. Furthermore, we apply the proposed approach to the National Health and Nutrition Examination Survey dataset to assess the impact of participation in school lunch programs on body mass index.

\vspace{0.5cm} \noindent{\bf KEY WORDS}: Double robust, Heterogeneous treatment effect, Single-index model, Sieve method
\vspace{3mm}

\let\thefootnote\relax\footnotetext{$^{\dag}$To whom correspondence should be addressed.}

\section{Introduction}

Observational studies are widely used to evaluate the treatment effect or policy effect
due to the fact that they usually contain rich information about markets, human beings and their behavior \citep{kunzel2019metalearners}. Traditional methods mainly focus on the average treatment effect of target population with observational studies.
With the development of technology and a large number of high-quality biomedical data being collected,
it is possible to conduct personalized medicine where the treatment  is customized for an individual patient \citep{anoke2019approaches}.
 In personalized medicine, it needs to evaluate the effectiveness of treatments tailored  to the covariate distribution of a
 population, which can be measured in terms of  {\it heterogeneous treatment effect} (HTE).
  If HTE exists, the population can be partitioned into subgroups across which the treatment effect differs.
  Therefore, HTE can help individuals make informed decisions on whether a particular treatment will work for
them.

Compared with the rich theory of average treatment effect (ATE), limited attention is paid to the estimation of HTE. Recently,
the estimation of HTE has become one of most popular topics in statistics and machine learning communities.
In the statistical community, \cite{crump2008nonparametric} proposed two
nonparametric tests for HTE with series estimation.
  \cite{tian2014simple} took  interactions between the treatment and covariates into account and used  the ordinary least square method to obtain the consistent estimator of  HTE. \cite{abrevaya2015estimating} used the kernel-based regression
to estimate HTE and established  established the corresponding theoretical properties.
\cite{lee2017doubly} proposed the double robust method to estimate HTE and established
a two-sided and symmetric uniform confidence band.
\cite{song2017semiparametric} estimated HTE with monotone B-splines under the framework of
the semiparametric additive single-index model
and established  the asymptotic properties of the proposed estimator.
\cite{feng2022statistical} considered the
 semiparametric single-index model to estimate HTE with
the refined minimum average variance estimation method.

In the machine learning community,  \cite{su2009subgroup} proposed
an interaction tree  procedure  to study HTE.
The Bayesian additive regression trees had been used to study
HTE by many authors (\citealp{green2012modeling,zeldow2019semiparametric,
hahn2020bayesian,hill2020bayesian}).
\cite{grimmer2017estimating} proposed  the ensemble methods
to consider the estimation of HTE.
The debiased machine learning methods had been proposed to study the estimation
of HTE (\citealp{chernozhukov2017double,semenova2021debiased}).
\cite{wager2018estimation} used the random forests to estimate HTE and established
the theoretical properties of the proposed estimator.
\cite{knaus2021machine} investigated the finite sample performance of machine learning methods to estimate HTE through an empirical Monte Carlo study.

Although machine learning methods are very powerful for prediction,
the results obtained by them are usually  too complicated to be interpretability,
which is very important in causal inference.
Besides, it is often difficult to draw valid inferences based on machine learning methods.
As is known to all, the parametric regression model can not capture  the complicated
relationship between the outcome and covariates.
The nonparametric regression model can be very flexible to capture the aforementioned relationship, which is  usually subject to the curse of dimensionality.
The semiparametric regression model
is a compromise
between the parametric regression model and nonparametric regression model,
which not only maintains the flexibility but also can
overcome the  curse of dimensionality. Hence, in this article, we consider
the semiparametric single-index model to study HTE.

The single-index model has been extensively studied in the statistics and econometrics literature,
with spline and kernel methods being the predominant approaches
(\citealp{xia2006asymptotic,yu2002penalized,ma2015varying}).
Among the most relevant works to our research are those by
\cite{song2017semiparametric} and \cite{feng2022statistical},
who employed the single-index model to estimate HTE in randomized experiments within the dynamic treatment regimes framework.
However, these studies neither adopt the counterfactual framework nor impose restrictions such as the boundedness of the link function or its support, thereby limiting their practical applicability. In contrast, our work investigates the single-index model for HTE estimation in observational studies under the counterfactual framework. Moreover, we leverage Hermite orthogonal polynomials to approximate the unknown link function, eliminating the need for boundedness assumptions on the link function or its support.

The  remainder  of the article is organized as follows. In Section 2, we introduce the notation,
 assumptions and  estimation procedures.
In Section 3, we establishes the theoretical  properties of the proposed estimators. In Section 4, simulation studies are conducted to assess the finite-sample performance of
the our methods.  Section 5  applies the proposed approach to a real dataset for illustration.
Finally, Section 6 provides concluding remarks and discussions,
while all technical proofs are deferred to the Appendix.

\section{Estimation of heterogeneous treatment effect}
\subsection{Notation, data structure and target estimand}
To fix  the notation, let $D$ denote the treatment assignment ($D=1$ for active treatment, $D=0$ for control) and $X=(X_1,\ldots, X_p)^{T}$ denote a vector of $p$-dimensional baseline
covariates (with $T$ indicating transpose).
Within the framework of
counterfactual (\citealp{rubin1974estimating,rubin1976inference}), we define $Y(1)$ and $Y(0)$
as the potential outcomes under treatment and control, respectively.
The fundamental problem of causal inference is that only one of potential outcomes
can be observed \citep{holland1986statistics}. The observed outcome $Y$ is equal to  $DY(1)+(1-D)Y(0)$.
 Hence, the observed data structure is $(Y,D,X)$.
 Assume $\{(Y_i,D_i,X_i), i=1,\ldots, n\}$ are independent copies of $(Y,D,X)$.
The average treatment effect is defined as
\begin{equation}\label{ATE}
\tau=E[Y(1)-Y(0)],
\end{equation}
a well-studied causal estimand in the literature.
The conditional average treatment effect (CATE) is given by
\begin{equation}\label{CATE}
\tau(x)=E[Y(1)-Y(0)|X=x].
\end{equation}
In this article, we focus on estimating HTE via CATE. Note that the ATE can be recovered by marginalizing CATE over the distribution $X$, i.e.,
$\tau=E[\tau(X)]$.

\subsection{Assumptions}
In order to identify  CATE, we have to make the following assumptions.

\begin{assumption}
(Stable unit treatment value assumption, SUTVA).
The treatment applied one individual does not affect other individuals' outcome
and  each individual has only one potential outcome under each treatment level.
\end{assumption}

\begin{assumption}
(No unmeasured confounding).  $(Y(1), Y(0))\bot D |X$, where the symbol $\bot$  indicates independence.
\end{assumption}

\begin{assumption}
(Overlap). For any $x\in\mathcal{X}$, there exits a positive  constant $0<\eta<1$  satisfying
$$\eta < P(D=1|X=x)< 1-\eta.$$
\end{assumption}

\begin{remark}
Assumption 2 means all the covariates that simultaneously affect the treatment assignment and
outcome  have been observed. Assumption 3 is to ensure the distribution of
covariates exist overlap without regard to which groups they come from.
\end{remark}

\begin{assumption} (Propensity score).
The treatment assignment is generated  by the following logistic regression model:
\begin{equation}\label{PS}
P(D=1|X)=\frac{\exp(\beta_{00}+\beta_{01}^{T}X)}{1+\exp(\beta_{00}+\beta_{01}^{T}X)},
\end{equation}
where $\beta_0=(\beta_{00},\beta_{01}^{T})^{T}$ is  in the space of
$\mathcal{R}^{p+1}$.
\end{assumption}

\begin{assumption} (Outcome model).
Given $d=\{0,1\}$, the parameter model $\mu_{d}(x;\alpha_{d,0})$  accurately represents the conditional expectation
$E[Y(d)|X=x]$.
\end{assumption}

\begin{remark}
The propensity score, introduced by \cite{rosenbaum1983central},  is defined as the conditional
probability of receiving an active  treatment,  given the observed baseline covariates.
The unknown parameter
$\beta_0$ can be estimated through the  maximum likelihood estimation. The unknown  parameters
$\alpha_{0,0}$ and $\alpha_{1,0}$  in the outcome models
 can be estimated by  the least squares method.

\end{remark}

\subsection{Estimation procedures}
Define $\mu_0(x)=E[Y(0)|X=x]$ and $\mu_1(x)=E[Y(1)|X=x]$. Then,
the conditional
average treatment effect $\tau(x)$ can be written as
\begin{eqnarray*}
\tau(x)=\mu_1(x)-\mu_0(x).
\end{eqnarray*}
Due to the fact that parametric models heavily depend on their assumptions and
nonparametric models are subject to the curse of dimensionality,
in this article,
we consider the semiparametric single-index model to estimate HTE,  which is given as follows
\begin{eqnarray}\label{SID}
Y=Dg_{0}(\gamma_{0}^{T}X)+\mu_0(X)+\epsilon,
\end{eqnarray}
where $Y$ is the observed outcome, $g_{0}(\cdot)$ is an unknown function, $\gamma_{0}$ is an unknown  $p$-dimensional regression parameters with $\|\gamma_0\|_2=1$ and $\gamma_{0,1}\geq 0$ for identifiability,  and the error term $\epsilon$ satisfies $E[\epsilon|X,D]=0$.
Under model \eqref{SID}, we can obtain
\begin{eqnarray}\label{CSID}
\tau(X)=g_{0}(\gamma_{0}^{T}X).
\end{eqnarray}

Let $\pi(X;\beta_0)$ and  $\pi(X;\widehat{\beta})$ denote the true and estimated propensity score, respectively. Under Assumption 2, we can have
\begin{eqnarray*}
\tau(X)=E\left[\frac{D(Y-\mu_{1}(X;\alpha_{1,0}))}{\pi(X;\beta_0)}-\frac{(1-D)(Y-\mu_{0}(X;\alpha_{0,0}))}
{1-\pi(X;\beta_0)}
+\mu_{1}(X;\alpha_{1,0})-\mu_{0}(X;\alpha_{0,0})|X\right].
\end{eqnarray*}
Hence, we
consider  the following semiparametric single-index model  to estimate HTE:
\begin{eqnarray*}\label{LADE}
(\widehat{\gamma},\widehat{g})&=&\arg\min_{\gamma,g,\lambda}
\sum\limits_{i=1}^{n}\left[
\frac{D_{i}(Y_{i}-\mu_{1}(X_i;\widehat{\alpha}_{1}))}{\pi(X_{i};\widehat{\beta})}-
\frac{(1-D_{i})(Y_{i}-\mu_{0}(X_i;\widehat{\alpha}_{0}))}{1-\pi(X_{i};\widehat{\beta})}\right.\\
&+&\left.\mu_{1}(X_i;\widehat{\alpha}_{1})-\mu_{0}(X_i;\widehat{\alpha}_{0})-
g(X_{i}^{T}\gamma)\right]^{2}
+\lambda\left(\|\gamma\|_{2}^{2}-1\right),
\end{eqnarray*}
where $\lambda $ is the Lagrange multiplier and $\|\cdot\|_{2}$ denotes the Euclidean norm.

Single-index models have gained considerable attention in the statistical literature
due to their flexibility in modeling complex relationships.
While nonparametric kernel and spline methods (\citealp{xia2006asymptotic, yu2002penalized, ma2015varying})
have been commonly employed for estimating single-index models, these approaches typically require restrictive boundedness assumptions on either the link function or its support. Such requirements may be violated in practical settings, particularly when estimated propensity scores approach the boundaries of zero or one. To overcome these limitations, we employ sieve methods, which offer both strong approximation capabilities for unknown functions and computational tractability \citep{Chen2007large}.
Specifically, we utilize Hermite orthogonal polynomials to approximate the unknown link function $g_{0}(\cdot)$, as this approach eliminates the need for boundedness assumptions on either the link function or its support.

Let $L^{2}(\mathbb{R},v(x))$ denote the Hilbert space with $v(x)=\exp(-x^2/2)$, which encompasses
 a broad class of functions including polynomials, power functions, and bounded functions
 on $\mathbb{R}$ (\citealp{dong2018additive,dong2019series}).
  We assume the unknown  link function
$g_{0}(\cdot)$ in model \eqref{SID} belongs to the Hilbert  space $L^{2}(\mathbb{R},v(x))$.
A commonly used standard orthogonal bases for $L^{2}(\mathbb{R},v(x))$ is provided by the Hermite orthogonal polynomials
$\{h_m(x),m=0,1,2,\ldots,\}$, defined as
\begin{eqnarray*}
h_m(x)=(\sqrt{2\pi}m!)^{-1/2}(-1)^{m}\exp(x^2/2)\frac{d^{m}}{dx^{m}}\exp(-x^2/2), m=0,1,2,\ldots.
\end{eqnarray*}
  For any function $g(x)\in L^{2}(\mathbb{R},v(x))$, it will have  an orthogonal series expansion in term of
$\{h_m(x),m=0,1,2,\ldots,\}$, which is given as follows
\begin{eqnarray*}
g(x)=\sum\limits_{m=0}^{\infty}c_mh_m(x)\quad \mbox{with}\quad c_m=\int g(x)h_m(x)v(x)dx.
\end{eqnarray*}
The  $L^2$-norm is given by
\begin{eqnarray*}
\|g\|_{L^{2}}=\left[\int |g(x)|^{2}v(x)dx\right]^{-1/2}=\left(\sum\nolimits_{m=0}^{\infty}c_{m}^{2}\right)^{-1/2},
\end{eqnarray*}
where the second equality follows from Parseval's identity.
 Hence, the function
$g(x)$ can be identified with its coefficient sequence $\{c_m, m=0,1,2,\ldots\}$ and
 regarded as a point in an infinite-dimensional Euclidean space. For any truncation parameter $k\geq 1$, the orthogonal series of
 $g(x)$ can be split into two parts:
 \begin{eqnarray}\label{f_link}
 g(x)=\sum\limits_{m=0}^{k-1}c_mh_m(x)+\sum\limits_{m=k}^{\infty}c_mh_m(x):=g_k(x)+\delta_{k}(x),
 \end{eqnarray}
where $g_k(x)=\sum\nolimits_{m=0}^{k-1}c_mh_m(x)$ and $\delta_{k}(x)=\sum\nolimits_{m=k}^{\infty}c_mh_m(x)$.
Under suitable regularity conditions,  $g_k(x) \rightarrow g(x)$ in $L^2$-norm (equivalently, $\delta_{k}(x)\rightarrow 0$)
as $k\rightarrow\infty$.

Define $\mathcal{H}(x)=(h_0(x),h_1(x),\ldots,h_{k-1}(x))^{T}$
 and  $\mathcal{C}_{k}=(c_0,c_1,\ldots,c_{k-1})^{T}$, then  $g_k(x)$ can be re-written as $g_k(x)=\mathcal{H}(x)^{T}\mathcal{C}_{k}$.
Due to  the expansion provided in \eqref{f_link}, the non-parametric function $g(\cdot)$ can be
parameterized with $\{c_m,m=0,1,\ldots\}$.
The non-parametric function $g(\cdot)$
and the unknown parametric $\gamma_0$
can be
 conceptualized as a single point
  in an infinite-dimensional
Euclidean space, which is the two-fold Cartesian product space by $L^{2}(\mathbb{R},v(\omega))$ and $\mathbb{R}^{p}$ and can be equipped with the norm
$\|\cdot\|$ being
\begin{eqnarray*}
\|(\gamma, g)\|_2=\left(\|\gamma\|_{2}^2+\sum\limits_{m=0}^{\infty}c_m^{2}\right)^{1/2}
\end{eqnarray*}
with $\|\gamma\|_2=\sqrt{\gamma_{1}^2+\cdots+\gamma_{p}^2}$.
Let
 $\Gamma$ be a compact subset of $\mathbb{R}^{p}$ with an interior  point denoted by $\gamma_0$  and $G$ be a subset of $L^{2}(\mathbb{R},v(\omega))$
 such that the supremum of the $L^2$-norm of all functions $g\in G$ is bounded by a large constant $M$. Let $g_0(\cdot)$ be an element of $G$
 and
define $G_k=G\cap \mbox{span}\{h_0(\omega),h_1(\omega),\ldots,h_{k-1}(\omega)\}$,
where $k$ represents the truncation parameter.

The estimators for $\gamma$ and $\mathcal{C}_{k}$ can be derived  by solving the constrained minimization problem:
\begin{eqnarray}\label{WLADE}
(\widehat{\gamma},\widehat{\mathcal{C}}_k)&=&\arg\min_{\Gamma,\Omega_{k},\lambda}
\sum\limits_{i=1}^{n}\left[
\frac{D_{i}(Y_{i}-\mu_{1}(X_{i},\widehat{\alpha}_{1}))}{\pi(X_{i};\widehat{\beta})}-\frac{(1-D_{i})(Y_{i}-\mu_{0}(X_{i},\widehat{\alpha}_{0}))}
{1-\pi(X_{i};\widehat{\beta})}\right.\nonumber\\
&+& \left.\mu_{1}(X_{i},\widehat{\alpha}_{1})-\mu_{0}(X_{i},\widehat{\alpha}_{0})-
\mathcal{H}(\gamma^{T}X_{i})^{T}\mathcal{C}_{k}\right]^{2}
+\lambda\left(\|\gamma\|_{2}^{2}-1\right),
\end{eqnarray}
where $\Omega_k=\{\mathcal{C}_k:\|\mathcal{C}_k\|_{2}\leq M\}\subset \mathbb{R}^{k}$.
\section{Asymptotic properties}
This section presents the asymptotic analysis of the proposed estimators.
 We begin by stating the required regularity conditions:

\begin{itemize}
\item[C1.] Let $\Gamma\subset \mathbb{R}^{p}$ be a convex and compact parameter space containing $\gamma_0$ as an interior point.
 The function space  $G$ is defined as:
 \begin{displaymath}
 G=\{g\in L^{2}(\mathbb{R},v(x)): \|g\|_{L^2}< M\},
 \end{displaymath}
 where $M>0$ is a sufficiently large constant bounding the $L^2$-norms uniformly.
 We assume the true link function $g_0\in G$.

\item[C2.] Define $\Sigma_1(\gamma_0,\mathcal{C}_{0,k})=E\left[(\mathcal{\dot{H}}
(X^{T}\gamma_0)^{T}\mathcal{C}_{0,k})^{2}XX^{T}\right]$, $\Sigma_2(\gamma_0)=E\left[\mathcal{\dot{H}}
(X^{T}\gamma_0)\mathcal{\dot{H}}
(X^{T}\gamma_0)^{T}\right]$,  and $\Sigma_3(\gamma_0)=E\left[\mathcal{H}
(X^{T}\gamma_0)\mathcal{H}(X^{T}\gamma_0)^{T}\right]$,
we have $M_1\leq\lambda_{min}(\Sigma_1(\gamma_0,\mathcal{C}_{0,k}))\leq
\lambda_{max}(\Sigma_1(\gamma_0,\mathcal{C}_{0,k}))\leq M_2$,
$M_3\leq\lambda_{min}(\Sigma_2(\gamma_0))\leq
\lambda_{max}(\Sigma_2(\gamma_0))\leq M_4$ and $M_5\leq\lambda_{min}(\Sigma_3(\gamma_0))\leq
\lambda_{max}(\Sigma_3(\gamma_0))\leq M_6$,
where $M_1$, $M_2$, $M_3$, $M_4$, $M_5$ and $M_6$ are positive constants,
 $\lambda_{max}(\Sigma)$ and  $\lambda_{min}(\Sigma)$ denote the maximum and
 minimum eigenvalues of matrix $\Sigma$.

\item[C3.] The link function $g_{0}(\cdot)$ is  $m$-order  differentiability on $\mathbb{R}$, where $m$ is a positive integer.
    Moreover,  each  $j$-order derivative  $g_{0}^{(j)}(\cdot)$,  for $j=0,1,\ldots, m$, belongs to the space $L^{2}(\mathbb{R},v(\omega))$.

\item[C4.] The truncation parameter $k$ is divergent with $n$ such that
$kn^{-1}\rightarrow 0$ and $nk^{-m}\rightarrow 0$ as $n \rightarrow \infty$,
where $m$ is defined in Condition C3.

\item[C5.] $E[\epsilon|D,X]=0$ and $E[\epsilon^{2}|D,X]=\sigma^{2}$, where $\sigma^{2}$
is a positive constant.

\item[C6.] $\sup\nolimits_{\{(\gamma, u)\in \Gamma\times \mathbb{R}\}}\exp(u^2/2)f_{\gamma}(u)\leq M$, where $f_{\gamma}(u)$
    is the probability density function of $u=X^{T}\gamma$ and
    $\sup\nolimits_{\{(\gamma, u)\in \Gamma\times G\}} E\|XX^{T}\{g^{(1)}(x^{T}\gamma)\}\|\leq M$.
\end{itemize}

\begin{remark}
Condition C1 provides the fundamental requirements for extremum estimation,  as highlighted by  Chen (2007).
Condition C2 adopts standard assumptions prevalent in single-index model literature (Yu and Ruppert, 2002).
The smoothness requirement in Condition C3 ensures proper control of truncation residuals, while Condition C4 governs their asymptotic behavior to facilitate normality results.
Condition C5 specifies an exogenous, homoscedastic error structure (though extensible to heteroscedastic cases), and finally, Condition C6 imposes moment restrictions to preclude heavy-tailed distributions-a crucial consideration given the potential unboundedness of the link function.
\end{remark}

\begin{theorem}\label{thm1}
Under Conditions C1-C6, we have
$\|(\widehat{\gamma},\widehat{g}_{k})-(\gamma_0,g_{0})\|_{2}\rightarrow_{p} 0$
as $n$ goes to infinity.
\end{theorem}
The proof of Theorem \ref{thm1} is given in  the Appendix.

Next, we will derive the asymptotic theory for the estimator $\widehat{\gamma}$.
Due to the the constraint $\|\gamma_0\|_{2}=1$, there is a projection matrix $P_{\gamma_0}=I_{p}-\gamma_{0}\gamma_{0}^{T}$,
which  maps any $p$-dimensional vector into the orthogonal complement space of the true parameter $\gamma_0$.
Consider the $p\times (p-1)$ matrix $V=[v_1,\ldots,v_{p-1}]$, where $\{v_j\}_{j=1}^{p-1}$
 are eigenvectors associated with the eigenvalue $1$ of the projection matrix $P_{\gamma_0}$.

\begin{theorem}\label{thm2}
Under Conditions C1-C6, we have
$\sqrt{n}V^{T}\left(\widehat{\gamma}-\gamma_0\right)\xrightarrow{d} N\left(0, \sigma^{2}(V^{T}\Sigma_{\Gamma} V)^{-1}\right)$
as $n$ goes to infinity, with the matrix $\Sigma_{\Gamma}$ being $E\left[\mathcal{C}_{0,k}^{T}\mathcal{\dot{H}}(\gamma_{0}^{T}X_{i})
\mathcal{\dot{H}}(\gamma_{0}^{T}X_{i})^{T}\mathcal{C}_{0,k}XX^{T}\right]$.
\end{theorem}
The proof of Theorem \ref{thm2} is given in  the Appendix.

In order to simplify the expression, define
$\Sigma_{A}=E\left[\mathcal{\dot{H}}(\gamma_{0}^{T}X)
\mathcal{\dot{H}}(\gamma_{0}^{T}X)^{T}\right]$,
$\Sigma_{B}=E\left[\mathcal{H}(\gamma_{0}^{T}X)\mathcal{H}(\gamma_{0}^{T}X)^{T}\right]$ and
$\Sigma_{D}=E\left[\mathcal{\dot{H}}(\gamma_0^{T}X)^{T}\mathcal{C}_{0,k}
\mathcal{H}(\gamma_{0}^{T}X)X^{T}\right]$.

\begin{theorem}\label{thm3}
Under Conditions C1-C6, we have
$
\sqrt{n}(\widehat{\mathcal{C}}_{k}-\mathcal{C}_{0,k})\xrightarrow{d}\mathcal{N}(0,\sigma^2\Sigma_{A}^{-1}\Sigma_{\mathcal{C}}\Sigma_{A}^{-1}),
$
as $n$ goes to infinity, with the matrix $\Sigma_{\mathcal{C}}$ being
 $\Sigma_{B}+
\Sigma_{D}V(V^{T}\Sigma_{\Gamma}V)^{-1}V^{T}\Sigma_{D}^{T}$.
\end{theorem}
The proof of Theorem \ref{thm3} is given in  the Appendix.

\section{Simulation studies}
In this Section, we conducted simulation studies to evaluate the finite-sample
performance of the proposed method. The response is generated by the following
model:
\begin{eqnarray}\label{Sim_m1}
Y=X_1-0.5X_2+
Dg_0(\gamma_1X_1+\gamma_2X_2+\gamma_3X_3)+\epsilon,
\end{eqnarray}
where $X_1$, $X_2$ and $X_3$ follow  the standard normal
distribution, $D$ denotes the treatment assignment, the link function  $g_0(x)=x$ or $g_0(x)=2x^3-1$,  and the random error $\epsilon$ follows
the standard normal distribution. The parameter
$(\gamma_1,\gamma_2,\gamma_3)$ is set to
$(0.8,-0.6,0)$. The treatment assignment
is generated by the following logistic regression model:
\begin{eqnarray}\label{Sim_m2}
P\left(D=1|X_1,X_2,X_3\right)=\frac{
\exp(\beta_0+X_1-0.5X_3)}
{1+\exp(\beta_0+X_1-0.5X_3)},
\end{eqnarray}
where $\beta_0$ is  chosen to yield different proportions of the treatment assignment.

We generate $1000$ simulated datasets, with the  sample size $n$ being set to $1000$ or $1500$.
The bias and sample standard deviation
of the $1000$ estimators for $(\gamma_1,\gamma_2,\gamma_3)$ are given in the columns ``Bias'' and ``SD'', respectively.
We employ $100$ bootstrap resamples to estimate the variance of the estimators.
The column ``ESE'' gives the average of the estimated standard errors and the column ``CI'' shows the nominal 95\% confidence interval coverage with the estimated
standard error.
The simulation results are summarized
in Table 1.
$$\mbox{Table 1 is about here}$$

From Table 1, we can make the following observations:
the proposed estimator remains approximately unbiased in all scenarios; the average standard error estimates closely match the sample standard deviations; and the confidence intervals achieve coverage rates near the nominal $95\%$ level.

We examine the scenario with a sample size of $n=1000$,
 where the treatment assignment proportion is set to $0.3$.
 The link function $g_{0}(x)$ is specified as either linear $g_0(x)=x$ or cubic $g_0(x)=2x^3-1$.
  The corresponding estimates of the link function are presented in Figure 1.
$$\mbox{Figure 1 is about here}$$

In Figure 1, the true link function is represented by the solid black line, while the estimated link function is depicted by the dashed red line. The shaded gray region indicates the $95\%$ confidence interval for the estimated function. As visually demonstrated, the estimated link function closely approximates the true link function.

We examine a scenario where the variables $X_1$, $X_2$ and $X_3$
are independently and identically distributed as uniform random variables on the interval $[-1,1]$
and the link function $g_{0}(x)$ is also specified as either linear $g_0(x)=x$ or cubic $g_0(x)=2x^3-1$.
The response variable is also generated according to Model \eqref{Sim_m1}, while the treatment assignment follows Model \eqref{Sim_m2}.
The sample size is set to $1000$ for all simulations.
We evaluate the performance of our proposed estimator $\tau_{SIM}$ by comparing it with three alternative approaches:
(1) $\tau_{CF}$, the causal forest estimator; (2) $\tau_{X-RF}$, the X-learner implemented with random forests; and (3)  $\tau_{T-RF}$, the T-learner based on random forests.
Figure 2 presents the simulation results, reporting the mean squared errors (MSEs) for all four methods.
$$\mbox{Figure 2 is about here}$$

As evident from Figure 2, the proposed method demonstrates superior performance across all considered scenarios. Specifically,
When the link function in the single-index model is linear, our method achieves comparable performance to both the causal forest and X-learner with random forest approaches, while the T-learner with random forest performs least favorably.
For the cubic link function case, our method shows significant advantages over all three alternatives, with the causal forest method exhibiting the poorest performance in this scenario.

To further assess the robustness of our proposed methodology,
we perform comprehensive simulation studies under various experimental conditions.
We systematically examine eight distinct model specification scenarios: the combination of misspecified propensity score, misspecified regression model for the response variable, and misspecified single-index model.
We assume
the response is generated by the following
model:
\begin{eqnarray}\label{Sim_m3}
Y=X_1-0.5X_2+
Dg_0(0.8X_1-0.6X_2)+\epsilon,
\end{eqnarray}
where $X_1$, $X_2$ and $X_3$ follow  the uniform
distribution over the interval $[-1,1]$, $D$ denotes the treatment assignment, the link function  $g_0(x)=x$,  and the random error $\epsilon$ follows
the standard normal distribution.
The treatment assignment is  generated by the following model:
\begin{eqnarray}\label{Sim_m4}
P\left(D=1|X_1,X_2,X_3\right)=\frac{
\exp(X_1-0.5X_3)}
{1+\exp(X_1-0.5X_3)}.
\end{eqnarray}
The proportion of the treatment assignment is equal to $0.5$.
We employ a Probit model to estimate the propensity scores that were originally specified using a Logistic model,
 indicating the presence of propensity score misspecification.
   We employ linear regression with quadratic terms ($X_{1}^2,X_{2}^2, X_{3}^2$)
 to detect misspecification in the originally specified linear model ($X_{1}, X_{2}, X_{3}$).
 Let $g(x)=g_0(x)-0.5x_3^{2}$ denote the misspecified form of $g_0(x)$.
Across all simulation settings, we set the sample size and treatment assignment proportion to $1000$ and $0.5$,
respectively. Figure 3 presents the simulation results, reporting the mean squared errors (MSEs) for the methods of
$\tau_{SIM}$,  $\tau_{CF}$, $\tau_{X-RF}$,  and $\tau_{T-RF}$.
We use an ordered triplet of binary indicators (T/F) to denote the misspecification status of the propensity score model, outcome  regression model, and single-index model, respectively.
$$\mbox{Figure 3 is about here}$$

From Figure 3, we demonstrate the robustness of our proposed method through three key observations:
 (1) when at least one of the propensity score or response regression models is correctly specified, our method outperforms all alternatives;
(2) under  misspecification of both models, our approach achieves comparable performance to $\tau_{CF}$ and $\tau_{X-RF}$ methods;
(3) the method maintains robustness against misspecification in the single-index model.
\section{National Health and Nutrition Examination Survey}
The National Health and Nutrition Examination Survey (NHANES) is designed to assess the health and nutritional status of children and adults in the United States. In this article,  we focuses on examining whether participation in the National School Lunch Program or School Breakfast Program leads to increased body mass index (BMI) among children and adolescents aged $4$ to $17$.
These meal programs aim to address food insecurity among children from low-income families but may have unintended consequences-excessive food intake could potentially contribute to childhood obesity.

Our analysis included a sample of $2,330$ children and adolescents aged $4$ to $17$ (median age: 10 years), of whom $1,284$ ($55\%$) participated in the school meal programs.
The outcome variable is the child's body mass index. The treatment variable, $school_{meal}$, is a binary indicator (1, eats meals at school; 0, otherwise). The  covariates include child characteristics ($age$ in years, $childsex$, race/ethnicity indicators [$afam$ for Black/African American and $hisam$ for Hispanic]), household socioeconomic factors ($povlev_{200}$ for income above $200\%$ of the poverty line, $supnut_{prog}$ and $foodstamp_{prog}$ for nutrition assistance participation, $foodsec_{chd}$ for food security, and $anyins$ for health insurance coverage), and respondent demographics ($refsex$ [1 = male] and $refage$ in years).
The parameter estimates and link function estimation in the single-index model are presented in Table 2 and Figure 4, respectively, while the estimates of individual treatment effects are shown in Figure 5.
$$\mbox{Table 2 is about here}$$
Based on the results presented in Table 2, we observe that among the $11$ covariates examined, only the afam variable (indicating Black/African American identity) shows statistical significance.
$$\mbox{Figure 4 is about here}$$
Figure 4 further reveals that the link function in the single-index model follows a linear form. The finding of only one significant covariate provides additional support for  a relatively simple relationship structure in the model.

We also evaluate the performance of our proposed estimator $\tau_{SIM}$ by comparing it with three alternative approaches:
(1) $\tau_{CF}$, the causal forest estimator; (2) $\tau_{X-RF}$, the X-learner implemented with random forests; and (3)  $\tau_{T-RF}$, the T-learner based on random forests.
Figure 5 presents the  results, reporting individual treatment effects for all four methods.
$$\mbox{Figure 5 is about here}$$
Figure 5 demonstrates that our method yields an average treatment effect (ATE) of zero, which aligns with existing research on this treatment effect. In contrast, alternative methods produce estimates that slightly deviate from zero.
Besides, the  results demonstrate comparable performance among the $\tau_{SIM}$,$\tau_{CF}$, and $\tau_{X-RF}$ methods, while $\tau_{T-RF}$ exhibits relatively inferior performance in comparison.

\section{Conclusion}
The study of heterogeneous treatment effects  addresses the limitations of traditional average treatment effects, enabling more refined and personalized causal inference. It holds significant practical value in fields such as medicine, economics, and policy evaluation. With the advancement of big data and machine learning, HTE analysis is becoming one of the core research directions in causal inference. In this article, to take advantages of semiparametric modeling, we employ single-index models to characterize heterogeneous treatment effects. Unlike conventional approaches, we neither impose boundedness assumptions on the link function nor require its support to be bounded in the single-index framework. Instead, we utilize the sieve method to approximate the link function, which enables simultaneous estimation of both the link function and index parameters. We establish the asymptotic properties of the proposed estimator and evaluate its finite-sample performance through comprehensive simulation studies. The numerical results demonstrate that our estimator outperforms other commonly used competing estimation methods.

This study adopts a single-index model framework for estimating heterogeneous treatment effects. The proposed estimation method demonstrates both favorable statistical properties and superior performance in practical applications. Future research could be extended along two promising directions: developing nonparametric estimation methods based on reproducing kernel Hilbert space theory, and  incorporating deep neural networks and other modern machine learning techniques to enhance estimation accuracy through their powerful feature extraction capabilities. These research directions offer significant theoretical value while holding substantial potential for practical applications.

In this study, we employ a logistic regression model to estimate propensity scores. It is noteworthy that \cite{Huang2025Semiparametric} proposed a single-index model for propensity score estimation. Building upon this foundation, we intend to further explore the application of single-index models in propensity score estimation in our future research.

\newpage
\section*{Appendix}

To establish the asymptotic properties of our proposed estimators,
we employ the following  lemma, which adapts results from Dong, Gao, and Peng (2019).

\begin{lemma}\label{lem1}
Assume $g^{(i)}(u)\in L^{2}(\mathbb{R},\pi(u))$ where $\pi(u)=\exp(-u^2/2)$, $0\leq i\leq m$. Then, $\|\delta_{k}(u)\|_{L^{2}}^{2}=O(k^{-m})$ and $\|\delta_{k}^{'}(u)\|_{L^{2}}^{2}=o(1)$ as $k\rightarrow \infty$.
\end{lemma}

{\bf Proof of Theorem 1}
To simplify the expression, define
\begin{eqnarray*}
L_n(\gamma,g)&=&\frac{1}{n}\sum\limits_{i=1}^{n}\left[
\frac{D_{i}(Y_{i}-\mu_{1}(X_i;\widehat{\alpha}_{1}))}{\pi(X_{i};\widehat{\beta})}
-\frac{(1-D_{i})(Y_{i}-\mu_{0}(X_i;\widehat{\alpha}_{0}))}{1-\pi(X_{i};\widehat{\beta})}\right.\\
&+&\left.\mu_{1}(X_i;\widehat{\alpha}_{1})-\mu_{0}(X_i;\widehat{\alpha}_{0})-
\mathcal{H}(\gamma^{T}X_{i})^{T}\mathcal{C}_{k}\right]^{2}
\end{eqnarray*}
and
 \begin{eqnarray*}
L(\gamma,g)&=&E\left[
\frac{D(Y-\mu_1(X;\alpha_{1,0}))}{\pi(X;\beta_0)}-\frac{(1-D)(Y-\mu_0(X;\alpha_{0,0}))}{1-\pi(X;\beta_0)}\right.\\
&+&\left.\mu_1(X;\alpha_{1,0})-\mu_0(X;\alpha_{0,0})
-g(\gamma^{T}X)\right]^{2}.
\end{eqnarray*}
To establish
the consistency of the estimators $(\widehat{\gamma},\widehat{g})$,
we first prove the uniform convergence:
\begin{eqnarray}\label{thm2_1}
\sup\nolimits_{\Gamma \times G}|L_n(\gamma,g)-L(\gamma,g)|\rightarrow_p 0.
\end{eqnarray}
Following
 Lemma A2 of Newey and Powell (2003), this convergence holds under the following  conditions:
  \begin{itemize}
\item[I.] $\Gamma \times G$ is a compact subset of a space with norm $\|(\gamma,g)\|_{2}$,
\item[II.] $L_n(\gamma,g)$ converges to $L(\gamma,g)$ in probability for all
$(\gamma,g)\in \Gamma \times G$,
\item [III.] $L_n(\gamma,g)$ is continuous for all $(\gamma,g)\in \Gamma \times G$.
\end{itemize}
  Condition I  follows directly from Condition C1.
 Condition II follows from the asymptotic equivalence
 $L_n(\gamma,g)=L(\gamma,g)(1+o_p(1))$,
 which holds under the weak law of large numbers and the consistency of the estimators
$\widehat{\beta}$, $\widehat{\alpha}_{1}$ and $\widehat{\alpha}_{0}$.
  Condition III  requires   $L_n(\gamma,g)$  is continuous.
  The term $L(\gamma,g)$ can be re-written as $L_{1}(\gamma,g)+\sigma^{2}$, where
   $L_{1}(\gamma,g)=E\left[g_{0}(X^{T}\gamma_0)-g(X^{T}\gamma)\right]^{2}$.
Subsequently,
we will demonstrate the continuity of
of $L_1(\gamma,g)$, which implies that
the continuity of $L_n(\gamma,g)$ holds with probability approaching one.

For any given $(\gamma_1,g_1)$ and $(\gamma_2,g_2) $ in $\Gamma \times G$,
we have
\begin{eqnarray}\label{thm2_2}
|L_1(\gamma_1,g_1)-L_1(\gamma_2,g_2)|\leq |L_1(\gamma_1,g_1)-L(\gamma_1,g_2)|+|L_1(\gamma_1,g_2)-
L_1(\gamma_2,g_2)|.
\end{eqnarray}
The first term of the right side of \eqref{thm2_2} is
\begin{eqnarray*}
|L(\gamma_1,g_1)-L(\gamma_1,g_2)|
&=&\left|E\left[(g_0(X^{T}\gamma_0)-g_1(X^{T}\gamma_1))^{2}
-(g_0(X^{T}\gamma_0)-g_2(X^{T}\gamma_1))^{2}\right]\right|\\
&=&\left|E\left[(g_2(X^{T}\gamma_1)-g_1(X^{T}\gamma_1))
(2g_0(X^{T}\gamma_0)-g_1(X^{T}\gamma_1)-g_2(X^{T}\gamma_1))\right]\right|\\
&\leq&\left\{E\left[g_2(X^{T}\gamma_1)-g_1(X^{T}\gamma_1)\right]^{2}E\left[2g_0(X^{T}\gamma_0)-g_1(X^{T}\gamma_1)-g_2(X^{T}\gamma_1)\right]^{2}\right\}^{1/2}
\end{eqnarray*}
by Cauchy-Schwarz inequality. Due to  Condition C6, we have
\begin{eqnarray*}
E[g_1(X^{T}\gamma_1)-g_2(X^{T}\gamma_1)]^{2}&=& \int(g_1(u)-g_2(u))^2f_{\gamma_1}(u)du\\
&=&\int(g_1(u)-g_2(u))^2\exp(-u^2/2)\exp(u^2/2)f_{\gamma_1}(u)du\\
&\leq & O(1)\int(g_1(u)-g_2(u))^2\exp(-u^2/2)du\\
&=&O(1)\|g_1-g_2\|_{L^2}^2 .
\end{eqnarray*}

For any $(\gamma,g)\in \Gamma \times G$, we can obtain
\begin{eqnarray*}
E[g(X^{T}\gamma)]^2&=&\int g^{2}(u)\exp(-u^2/2)\exp(u^2/2)f_{\gamma}(u)du\\
&\leq & O(1)\|g\|_{L^2}^2\leq O(1),
\end{eqnarray*}
where  the boundedness is guaranteed by Condition C6 and the compactness of the set $\Gamma \times G$.
Similarly, we have
\begin{eqnarray*}
E\left[2g_0(X^{T}\gamma_0)-g_1(X^{T}\gamma_1)-g_2(X^{T}\gamma_1)\right]^{2}&\leq&8E\left[g_0(X^{T}\gamma_{0})\right]^{2}
+4E\left[g_1(X^{T}\gamma_{1})\right]^{2}\\
&+&4E\left[g_2(X^{T}\gamma_{1})\right]^{2}
\leq O(1).
\end{eqnarray*}
Hence, we  have
\begin{eqnarray*}\label{thm2_3}
|L_1(\gamma_1,g_1)-L_1(\gamma_1,g_2)|\leq O(1)\|g_1-g_2\|_{L^2}.
\end{eqnarray*}

The second term of the right side of \eqref{thm2_2} satisfies
\begin{eqnarray*}
	|L(\gamma_1,g_2)-L(\gamma_2,g_2)|
	&=&\left|E\left[(g_0(X^{T}\gamma_0)-g_2(X^{T}\gamma_1))^{2}
	-(g_0(X^{T}\gamma_0)-g_2(X^{T}\gamma_2))^{2}\right]\right|\\
	&=&\left|E\left[(g_2(X^{T}\gamma_2)-g_2(X^{T}\gamma_1))
	(2g_0(X^{T}\gamma_0)-g_2(X^{T}\gamma_1)-g_2(X^{T}\gamma_2))\right]\right|\\
	&\leq&\left\{E\left[g_2(X^{T}\gamma_2)-g_2(X^{T}\gamma_1)\right]^{2}E\left[2g_0(X^{T}\gamma_0)-g_2(X^{T}\gamma_1)-g_2(X^{T}\gamma_2)\right]^{2}\right\}^{1/2}.
\end{eqnarray*}
Similar to the above, it is easy to show the term $E\left[2g_0(X^{T}\gamma_0)-g_2(X^{T}\gamma_1)-g_2(X^{T}\gamma_2)\right]^{2}$ is bounded uniformly on $\Gamma \times G$.
The
term $E\left[g_2(X^{T}\gamma_2)-g_2(X^{T}\gamma_1)\right]^{2}$
can be rewritten as
\begin{eqnarray*}
E\left[g_2(X^{T}\gamma_2)-g_2(X^{T}\gamma_1)\right]^{2}
&=&E\left[(\gamma_2-\gamma_1)^{T}XX^{T}(\gamma_2-\gamma_1)
\{g_{2}^{(1)}(X^{T}\gamma^{*})\}^{2}\right]\\
&\leq&\|\gamma_2-\gamma_1\|_{2}^{2}E\|XX^{T}\{g_{2}^{(1)}(X^{T}\gamma^{*})\}^{2}\|\\
&\leq&O(1)\|\gamma_2-\gamma_1\|_{2}^{2},
\end{eqnarray*}
where $\gamma^{*}$ lies between $\gamma_1$ and $\gamma_2$.
Hence, we have
\begin{eqnarray*}
|L_{1}(\gamma_1,g_2)-L_{1}(\gamma_2,g_2)|\leq O(1)\|\gamma_2-\gamma_1\|_{2}.
\end{eqnarray*}
Furthermore, we obtain
\begin{eqnarray*}
|L(\gamma_1,g_1)-L(\gamma_2,g_2)|\leq O(1)\|(\gamma_1,g_1)-
(\gamma_2,g_2)\|,
\end{eqnarray*}
which indicates the continuity of $L(\gamma,g)$.
Then, we can obtain
\begin{eqnarray*}
\sup\nolimits_{\Gamma \times G}|L_n(\gamma,g)-L(\gamma,g)|\rightarrow_p 0
\end{eqnarray*}
by Lemma A2 of Newey and Powell (2003).  Then,
If the estimator $(\widehat{\gamma},\widehat{g})$ is not  consistent for  $(\gamma_0,g_{0})$,  which will destroy the definition of \eqref{WLADE}.
Therefore, Theorem \ref{thm1} holds.

{\bf Proof of Theorem 2}\quad
To streamline the expression, we define
\begin{eqnarray*}
\widehat{Y}_i=\frac{D_{i}(Y_{i}-\mu_{1}(X_i;\widehat{\alpha}_{1}))}
{\pi(X_{i};\widehat{\beta})}-\frac{(1-D_{i})(Y_{i}-\mu_{0}(X_i;\widehat{\alpha}_{0}))}{1-\pi(X_{i};\widehat{\beta})}
+\mu_{1}(X_i;\widehat{\alpha}_{1})-\mu_{0}(X_i;\widehat{\alpha}_{0}),
\end{eqnarray*}
\begin{eqnarray*}
Y_{i}^{*}=\frac{D_{i}(Y_{i}-\mu_{1}(X_i;\alpha_{1,0}))}
{\pi(X_{i};\beta_0)}-\frac{(1-D_{i})(Y_{i}-\mu_{0}(X_i;\alpha_{0,0}))}{1-\pi(X_{i};\beta_0)}
+\mu_{1}(X_i;\alpha_{1,0})-\mu_{0}(X_i;\alpha_{0,0})
\end{eqnarray*}
and
 \begin{eqnarray*}
L_{n}(\gamma,\mathcal{C}_{k},\lambda)&=&\frac{1}{n}\sum\limits_{i=1}^{n}\left[
\frac{D_{i}(Y_{i}-\mu_{1}(X_i;\widehat{\alpha}_{1}))}{\pi(X_{i};\widehat{\beta})}-\frac{(1-D_{i})(Y_{i}-\mu_{0}(X_i;\widehat{\alpha}_{0}))}
{1-\pi(X_{i};\widehat{\beta})}\right.\\
&+&\left. \mu_{1}(X_i;\widehat{\alpha}_{1})-\mu_{0}(X_i;\widehat{\alpha}_{0})-
\mathcal{H}(\gamma^{T}X_{i})^{T}\mathcal{C}_{k}\right]^{2}+\lambda(\|\gamma\|_{2}^{2}-1)\\
&=&\frac{1}{n}\sum\limits_{i=1}^{n}\left[\widehat{Y}_{i}-\mathcal{H}(\gamma^{T}X_{i})^{T}\mathcal{C}_{k}\right]^{2}
+\lambda(\|\gamma\|_{2}^{2}-1).
\end{eqnarray*}

Let $U_{n,1}(\gamma,\mathcal{C}_{k},\lambda)$, $U_{n,2}(\gamma,\mathcal{C}_{k},\lambda)$ and $U_{n,3}(\gamma,\mathcal{C}_{k},\lambda)$ denote
$\partial L_{n}(\gamma,\mathcal{C}_{k},\lambda)/\partial \gamma$,
$\partial L_{n}(\gamma,\mathcal{C}_{k},\lambda)/\partial \mathcal{C}_{k}$ and
$\partial L_{n}(\gamma,\mathcal{C}_{k},\lambda)/\partial \lambda$, respectively, where
\begin{eqnarray*}
U_{n,1}(\gamma,\mathcal{C}_{k},\lambda)&=&-\frac{2}{n}\sum_{i=1}^{n}
\left[\widehat{Y}_{i}-\mathcal{H}(\gamma^{T}X_{i})^{T}\mathcal{C}_{k}\right]\mathcal{\dot{H}}(\gamma^{T}X_{i})^{T}\mathcal{C}_{k}X_i+2\lambda\gamma,
\end{eqnarray*}
\begin{eqnarray*}
	U_{n,2}(\gamma,\mathcal{C}_{k},\lambda)&=&-\frac{2}{n}\sum_{i=1}^{n}
	\left[\widehat{Y}_{i}-\mathcal{H}(\gamma^{T}X_{i})^{T}\mathcal{C}_{k}\right]\mathcal{H}(\gamma^{T}X_{i}),
\end{eqnarray*}
and
\begin{eqnarray*}
	U_{n,3}(\gamma,\mathcal{C}_{k},\lambda)&=&\|\gamma\|_{2}^{2}-1.
\end{eqnarray*}
By multiplying $\widehat{\gamma}$ in both sides of $\partial L_{n}(\widehat{\gamma},\widehat{\mathcal{C}}_{k},\widehat{\lambda})/\partial
\gamma=0$, we can obtain:
\begin{eqnarray*}
\widehat{\lambda}=\frac{1}{n}\sum_{i=1}^{n}
\left[\widehat{Y}_{i}-\mathcal{H}(\widehat{\gamma}^{T}X_{i})^{T}\widehat{\mathcal{C}}_{k}\right]
\mathcal{\dot{H}}(\widehat{\gamma}^{T}X_{i})^{T}\widehat{\mathcal{C}}_{k}X_{i}^{T}\widehat{\gamma}.
\end{eqnarray*}

Next, we will focus on the term $\frac{\partial U_{n,1}(\gamma,\mathcal{C}_{k},\lambda) }
{\partial \gamma}|_{(\gamma,\mathcal{C}_{k},\lambda)=(\bar{\gamma},\widehat{\mathcal{C}}_{k},\widehat{\lambda})}$
with $\bar{\gamma}$ lies between $\widehat{\gamma}$ and $\gamma_0$, where
\begin{eqnarray*}
\frac{\partial U_{n,1}(\gamma,\mathcal{C}_{k},\lambda) }{\partial \gamma}	&=&\frac{2}{n}\sum_{i=1}^{n}
\mathcal{C}_{k}^{T}\mathcal{\dot{H}}(\gamma^{T}X_{i})
\mathcal{\dot{H}}(\gamma^{T}X_{i})^{T}\mathcal{C}_{k}
X_{i}X_{i}^{T}\\
&-&\frac{2}{n}\sum_{i=1}^{n}
\left[\widehat{Y}_{i}-\mathcal{H}(\gamma^{T}X_{i})^{T}\mathcal{C}_{k}\right]
\mathcal{\ddot{H}}(\gamma^{T}X_{i})^{T}\mathcal{C}_{k}
X_{i}X_{i}^{T}+2\lambda.
\end{eqnarray*}
In order to simplify the expression, we define
\begin{eqnarray*}
\frac{\partial U_{n,1}(\gamma,\mathcal{C}_{k},\lambda) }
{\partial \gamma}|_{(\gamma,\mathcal{C}_{k},\lambda)=(\bar{\gamma},\widehat{\mathcal{C}}_{k},\widehat{\lambda})}
=2J_{n,1}|_{(\gamma,\mathcal{C}_{k})=(\bar{\gamma},\widehat{\mathcal{C}}_{k})}-2
J_{n,2}|_{(\gamma,\mathcal{C}_{k})=(\bar{\gamma},\widehat{\mathcal{C}}_{k})}+2\lambda I_{p}|_{\lambda=\widehat{\lambda}},
\end{eqnarray*}
where the definitions of $J_{n,1}$ and $J_{n,2}$ should be obvious.

Under Theorem 1 and by the law of large numbers, we have
\begin{eqnarray*}
J_{n,1}&=&\frac{1}{n}\sum_{i=1}^{n}
\widehat{\mathcal{C}}_{k}^{T}\mathcal{\dot{H}}(\bar{\gamma}^{T}X_{i})
\mathcal{\dot{H}}(\bar{\gamma}^{T}X_{i})^{T}\widehat{\mathcal{C}}_{k}
X_{i}X_{i}^{T}\\
&=&\frac{1}{n}\sum_{i=1}^{n}
\mathcal{C}_{0,k}^{T}\mathcal{\dot{H}}(\bar{\gamma}^{T}X_{i})
\mathcal{\dot{H}}(\bar{\gamma}^{T}X_{i})^{T}\mathcal{C}_{0,k}
X_{i}X_{i}^{T}\\
&+&\frac{1}{n}\sum_{i=1}^{n}
(\widehat{\mathcal{C}}_{k}-\mathcal{C}_{0,k})^{T}\mathcal{\dot{H}}(\bar{\gamma}^{T}X_{i})
\mathcal{\dot{H}}(\bar{\gamma}^{T}X_{i})^{T}(\widehat{\mathcal{C}}_{k}-\mathcal{C}_{0,k})
X_{i}X_{i}^{T}\\
&+&\frac{2}{n}\sum_{i=1}^{n}
(\widehat{\mathcal{C}}_{k}-\mathcal{C}_{0,k})^{T}\mathcal{\dot{H}}(\bar{\gamma}^{T}X_{i})
\mathcal{\dot{H}}(\bar{\gamma}^{T}X_{i})^{T}\mathcal{C}_{0,k}
X_{i}X_{i}^{T}\\
&=& J_{n,11}+J_{n,12}+2J_{n,13},
\end{eqnarray*}
where the definitions of $J_{n,11}$, $J_{n,12}$ and $J_{n,13}$ should be obvious.
For  $J_{n,12}$, we have
\begin{eqnarray*}
\|J_{n,12}\|&\leq &\left\|E[(\widehat{\mathcal{C}}_{k}-\mathcal{C}_{0,k})^{T}\mathcal{\dot{H}}(\bar{\gamma}^{T}X)
\mathcal{\dot{H}}(\bar{\gamma}^{T}X)^{T}(\widehat{\mathcal{C}}_{k}-\mathcal{C}_{0,k})XX^{T}]\right\|+o_p(1)\\
&\leq &\left\{E|(\widehat{\mathcal{C}}_{k}-\mathcal{C}_{0,k})^{T}\mathcal{\dot{H}}(\bar{\gamma}^{T}X)|^{4}E\|XX^{T}\|^{2}\right\}^{1/2}+o_p(1)\\
&\leq &O(1)\|\widehat{\mathcal{C}}_{k}-\mathcal{C}_{0,k}\|_{2}^{2}+o_p(1)=o_p(1).
\end{eqnarray*}

For $J_{n,13}$, we have
\begin{eqnarray*}
\|J_{n,13}\|&\leq &\left\|E[(\widehat{\mathcal{C}}_{k}-\mathcal{C}_{0,k})^{T}\mathcal{\dot{H}}(\bar{\gamma}^{T}X)
\mathcal{\dot{H}}(\bar{\gamma}^{T}X)^{T}\mathcal{C}_{0,k}XX^{T}]\right\|+o_p(1)\\
&\leq &\left\{E|(\widehat{\mathcal{C}}_{k}-\mathcal{C}_{0,k})^{T}\mathcal{\dot{H}}(\bar{\gamma}^{T}X)|^{2}E\|\mathcal{\dot{H}}(\bar{\gamma}^{T}X)^{T}
\mathcal{C}_{0,k}XX^{T}\|^{2}\right\}^{1/2}+o_p(1)\\
&\leq &O(1)\|\widehat{\mathcal{C}}_{k}-\mathcal{C}_{0,k}\|_{2}+o_p(1)=o_p(1).
\end{eqnarray*}
Hence, $J_{n,1}=E\left[\mathcal{C}_{0,k}^{T}\mathcal{\dot{H}}(\gamma_{0}^{T}X)
\mathcal{\dot{H}}(\gamma_{0}^{T}X)^{T}\mathcal{C}_{0,k}
XX^{T}\right]+o_p(1)$.

The term $J_{n,2}$ can be re-written as
\begin{eqnarray*}
J_{n,2}&=&\frac{1}{n}\sum_{i=1}^{n}
\left[\mathcal{H}(\gamma_{0}^{T}X_{i})^{T}\mathcal{C}_{0,k}-\mathcal{H}(\bar{\gamma}^{T}X_{i})^{T}\widehat{\mathcal{C}}_{k}\right]
\mathcal{\ddot{H}}(\bar{\gamma}^{T}X_{i})^{T}\widehat{\mathcal{C}}_{k}
X_{i}X_{i}^{T}\\
&+&\frac{1}{n}\sum_{i=1}^{n}
\delta_{0,k}(\gamma_0^{T}X_{i})
\mathcal{\ddot{H}}(\bar{\gamma}^{T}X_{i})^{T}\widehat{\mathcal{C}}_{k}
X_{i}X_{i}^{T}\\
&+&\frac{1}{n}\sum_{i=1}^{n}
\left(\widehat{Y}_i-Y_{i}^{*}\right)
\mathcal{\ddot{H}}(\bar{\gamma}^{T}X_{i})^{T}\widehat{\mathcal{C}}_{k}
X_{i}X_{i}^{T}\\
&=&J_{n,21}+J_{n,22}+J_{n,23},
\end{eqnarray*}
where the definitions of $J_{n,21}$, $J_{n,22}$ and $J_{n,23}$ should be obvious.

The term $J_{n,21}$ can be re-written as
\begin{eqnarray*}
J_{n,21}&=&\frac{1}{n}\sum_{i=1}^{n}
\left[\mathcal{H}(\gamma_{0}^{T}X_{i})^{T}\mathcal{C}_{0,k}-\mathcal{H}
(\bar{\gamma}^{T}X_{i})^{T}\mathcal{C}_{0,k}\right]
\mathcal{\ddot{H}}(\bar{\gamma}^{T}X_{i})^{T}(\widehat{\mathcal{C}}_{k}-\mathcal{C}_{0,k})X_{i}X_{i}^{T}\\
&+&\frac{1}{n}\sum_{i=1}^{n}
\left[\mathcal{H}(\bar{\gamma}^{T}X_{i})^{T}\mathcal{C}_{0,k}-\mathcal{H}
(\bar{\gamma}^{T}X_{i})^{T}\widehat{\mathcal{C}}_{k}\right]
\mathcal{\ddot{H}}(\bar{\gamma}^{T}X_{i})^{T}(\widehat{\mathcal{C}}_{k}-\mathcal{C}_{0,k})X_{i}X_{i}^{T}\\
&+&\frac{1}{n}\sum_{i=1}^{n}
\left[\mathcal{H}
(\gamma_{0}^{T}X_{i})^{T}\mathcal{C}_{0,k}-\mathcal{H}(\bar{\gamma}^{T}X_{i})^{T}\mathcal{C}_{0,k}\right]
\mathcal{\ddot{H}}(\bar{\gamma}^{T}X_{i})^{T}\mathcal{C}_{0,k}X_{i}X_{i}^{T}\\
&+&\frac{1}{n}\sum_{i=1}^{n}
\left[\mathcal{H}
(\bar{\gamma}^{T}X_{i})^{T}\mathcal{C}_{0,k}-\mathcal{H}(\bar{\gamma}^{T}X_{i})^{T}\widehat{\mathcal{C}}_{k}\right]
\mathcal{\ddot{H}}(\bar{\gamma}^{T}X_{i})^{T}\mathcal{C}_{0,k}X_{i}X_{i}^{T}\\
&=&J_{n,211}+J_{n,212}+J_{n,213}+J_{n,214},
\end{eqnarray*}
where the definitions of $J_{n,211}$, $J_{n,212}$, $J_{n,213}$  and $J_{n,214}$ should be obvious.
For $J_{n,211}$, we have
\begin{eqnarray*}
\|J_{n,211}\|&\leq &\left\|E[ (\gamma_0-\bar{\gamma})^{T}X\mathcal{\dot{H}}(\gamma_{*}^{T}X)^{T}\mathcal{C}_{0,k}
\mathcal{\ddot{H}}(\bar{\gamma}^{T}X)^{T}(\widehat{\mathcal{C}}_{k}-\mathcal{C}_{0,k})XX^{T}] \right\|+o_p(1)\\
&\leq& \left\{E|(\gamma_0-\bar{\gamma})^{T}X\mathcal{\dot{H}}(\gamma_{*}^{T}X)^{T}\mathcal{C}_{0,k}
\mathcal{\ddot{H}}(\bar{\gamma}^{T}X)^{T}(\widehat{\mathcal{C}}_{k}-\mathcal{C}_{0,k})|^{2}E\|XX^{T}\|^{2}\right\}^{1/2}+o_p(1)\\
&\leq & \left\{E|(\gamma_0-\bar{\gamma})^{T}X\mathcal{\dot{H}}(\gamma_{*}^{T}X)^{T}\mathcal{C}_{0,k}|^{2}
E|\mathcal{\ddot{H}}(\bar{\gamma}^{T}X)^{T}(\widehat{\mathcal{C}}_{k}-\mathcal{C}_{0,k})|^{2}\right\}^{1/2}E\|XX^{T}\|+o_p(1)\\
&\leq & O(1) \left\{ (\gamma_0-\bar{\gamma})^{T}E[X\mathcal{\dot{H}}(\gamma_{*}^{T}X)^{T}
\mathcal{C}_{0,k}\mathcal{C}_{0,k}^{T}\mathcal{\dot{H}}(\gamma_{*}^{T}X)X^{T}](\gamma_0-\bar{\gamma})\right\}^{1/2}\\
&\times&\left\{ (\widehat{\mathcal{C}}_{k}-\mathcal{C}_{0,k})^{T}E[\mathcal{\ddot{H}}(\bar{\gamma}^{T}X)\mathcal{\ddot{H}}(\bar{\gamma}^{T}X)^{T}]
(\widehat{\mathcal{C}}_{k}-\mathcal{C}_{0,k})\right\}^{1/2}+o_p(1)\\
&\leq&O(1)(\|\widehat{\gamma}-\gamma_0\|_{2}^{2}+\|\widehat{\mathcal{C}}_{k}-\mathcal{C}_{0,k}\|_{2}^{2})+o_p(1)=o_p(1).
\end{eqnarray*}
Similarly, we obtain $J_{n,212}=o_p(1)$, $J_{n,213}=o_p(1)$ and $J_{n,214}=o_p(1)$. Hence, $J_{n,21}=o_p(1)$.

The term $J_{n,22}$ can be re-written as:
\begin{eqnarray*}
J_{n,22}&=&\frac{1}{n}\sum_{i=1}^{n}
\delta_{0,k}(\gamma_0^{T}X_{i})
\mathcal{\ddot{H}}(\bar{\gamma}^{T}X_{i})^{T}(\widehat{\mathcal{C}}_{k}-\mathcal{C}_{0,k})
X_{i}X_{i}^{T}\\
&+&\frac{1}{n}\sum_{i=1}^{n}
\delta_{0,k}(\gamma_0^{T}X_{i})
\mathcal{\ddot{H}}(\bar{\gamma}^{T}X_{i})^{T}\mathcal{C}_{0,k}
X_{i}X_{i}^{T}\\
&=&J_{n,221}+J_{n,222},
\end{eqnarray*}
where  the definitions of $J_{n,221}$ and $J_{n,222}$ should be obvious.
For $J_{n,221}$, we have
\begin{eqnarray*}
\|J_{n,221}\|&\leq& O(1) \left\{(\widehat{\mathcal{C}}_{k}-\mathcal{C}_{0,k})^{T}
E[\mathcal{\ddot{H}}(\bar{\gamma}^{T}X_{i})\mathcal{\ddot{H}}(\bar{\gamma}^{T}X_{i})^{T}](\widehat{\mathcal{C}}_{k}-\mathcal{C}_{0,k})^{T}
\right\}^{1/2}+o_p(1)\\
&\leq&O(1)\|\widehat{\mathcal{C}}_{k}-\mathcal{C}_{0,k}\|_2+o_p(1)=o_p(1).
\end{eqnarray*}
For $J_{n,222}$, we have
\begin{eqnarray*}
\|J_{n,222}\|&\leq& \left\|E[\delta_{0,k}(\gamma_0^{T}X)
\mathcal{\ddot{H}}(\bar{\gamma}^{T}X)^{T}\mathcal{C}_{0,k}
XX^{T}]\right\|+o_p(1)\\
&\leq &O(1)\left[E|\delta_{0,k}(\gamma_0^{T}X)|^{2}\right]^{1/2}+o_p(1)=o_p(1).
\end{eqnarray*}
Hence, $J_{n,22}=o_p(1)$.

For   $J_{n,23}$, we have
\begin{eqnarray*}
J_{n,23}&=&	\frac{1}{n}\sum_{i=1}^{n}
\left(\widehat{Y}_i-Y_{i}^{*}\right)
\mathcal{\ddot{H}}(\bar{\gamma}^{T}X_{i})^{T}\widehat{\mathcal{C}}_{k}
X_{i}X_{i}^{T}\\
&=&-\frac{1}{n}\sum_{i=1}^{n}
\left[\frac{D_i(Y_i-\mu_{1}(X_i;\alpha_{1,0}))}{\pi^{2}(X_i;\beta^{*})}
\right]
\mathcal{\ddot{H}}(\bar{\gamma}^{T}X_{i})^{T}\widehat{\mathcal{C}}_{k}
X_{i}X_{i}^{T}\pi^{(1)}(X_i;\beta^{*})^{T}(\widehat{\beta}-\beta_0)\\
&+&\frac{1}{n}\sum_{i=1}^{n}\left[1-\frac{D_i}{\pi(X_i;\widehat{\beta})}\right]
\mathcal{\ddot{H}}(\bar{\gamma}^{T}X_{i})^{T}\widehat{\mathcal{C}}_{k}
X_{i}X_{i}^{T}\mu_{1}^{(1)}(X_i;\alpha_{1}^{*})^{T}(\widehat{\alpha}_1-\alpha_{1,0})\\
&-&\frac{1}{n}\sum_{i=1}^{n}
\left[\frac{(1-D_i)(Y_i-\mu_{0}(X_i;\alpha_{0,0}))}{(1-\pi(X_i;\beta^{*}))^{2}}
\right]
\mathcal{\ddot{H}}(\bar{\gamma}^{T}X_{i})^{T}\widehat{\mathcal{C}}_{k}
X_{i}X_{i}^{T}\pi^{(1)}(X_i;\beta^{*})^{T}(\widehat{\beta}-\beta_0)\\
&+&\frac{1}{n}\sum_{i=1}^{n}\left[\frac{1-D_i}{1-\pi(X_i;\widehat{\beta})}-1\right]
\mathcal{\ddot{H}}(\bar{\gamma}^{T}X_{i})^{T}\widehat{\mathcal{C}}_{k}
X_{i}X_{i}^{T}\mu_{0}^{(1)}(X_i;\alpha_{0}^{*})^{T}(\widehat{\alpha}_0-\alpha_{0,0})\\
&=&o_p(1),
\end{eqnarray*}
where
$\beta^{*}\in (\widehat{\beta}, \beta_0)$, $\alpha_{1}^{*}\in (\widehat{\alpha}_{1}, \alpha_{1,0})$ and  $\alpha_{0}^{*}\in (\widehat{\alpha}_{0}, \alpha_{0,0})$ are intermediate values. Hence, $J_{n,23}=o_p(1)$. Furthermore, $J_{n,2}=o_p(1)$.

Similarly, we can show $\widehat{\lambda}=o_p(1)$.
Therefore,  we have concluded $\frac{\partial U_{n,1}(\gamma,\mathcal{C}_{k},\lambda) }
{\partial \gamma}|_{(\gamma,\mathcal{C}_{k},\lambda)=(\bar{\gamma},\widehat{\mathcal{C}}_{k},\widehat{\lambda})}
=2E\left[\mathcal{C}_{0,k}^{T}\mathcal{\dot{H}}(\gamma_{0}^{T}X)
\mathcal{\dot{H}}(\gamma_{0}^{T}X)^{T}\mathcal{C}_{0,k}
XX^{T}\right]+o_p(1)$.

Expand $\sqrt{n} U_{n1}(\gamma,\mathcal{C}_{k},\lambda)|_{(\gamma,\mathcal{C}_{k},\lambda)
=(\gamma_0,\widehat{\mathcal{C}}_{k},\widehat{\lambda})}$ as follows:
\begin{eqnarray*}
\sqrt{n} U_{n1}(\gamma_0,\widehat{\mathcal{C}}_{k},\widehat{\lambda})
&=&-\frac{2}{\sqrt{n}}\sum_{i=1}^{n}
\left[\widehat{Y}_{i}-\mathcal{H}(\gamma_0^{T}X_{i})^{T}\widehat{\mathcal{C}}_{k}\right]
\mathcal{\dot{H}}(\gamma_0^{T}X_{i})^{T}\widehat{\mathcal{C}}_{k}X_i+2\sqrt{n}\widehat{\lambda}\gamma_0\\
&=&-\frac{2}{\sqrt{n}}\sum_{i=1}^{n}
\left[Y_{i}^{*}-\mathcal{H}(\gamma_0^{T}X_{i})^{T}\widehat{\mathcal{C}}_{k}\right]
\mathcal{\dot{H}}(\gamma_{0}^{T}X_{i})^{T}\widehat{\mathcal{C}}_{k}
X_i
\\
&-&\frac{2}{\sqrt{n}}\sum_{i=1}^{n}
\left(\widehat{Y}_{i}-Y_{i}^{*}\right)\mathcal{\dot{H}}(\gamma_{0}^{T}X_{i})^{T}\mathcal{C}_{0,k}X_i+
2\sqrt{n}\widehat{\lambda}\gamma_0\\
&=&-2\bar{J}_{n,1}-2\bar{J}_{n,2}+2\bar{J}_{n,3},
\end{eqnarray*}
where the definitions of $\bar{J}_{n,1}$, $\bar{J}_{n,2}$ and
$\bar{J}_{n, 3}$ are obvious.

For $\bar{J}_{n,1}$, we have
\begin{eqnarray*}
\bar{J}_{n,1}&=&\frac{1}{\sqrt{n}}\sum_{i=1}^{n}
\left[Y_{i}^{*}-\mathcal{H}(\gamma_0^{T}X_{i})^{T}\widehat{\mathcal{C}}_{k}\right]
\mathcal{\dot{H}}(\gamma_{0}^{T}X_{i})^{T}\widehat{\mathcal{C}}_{k}X_i\\
&=&
\frac{1}{\sqrt{n}}\sum_{i=1}^{n}
\mathcal{\dot{H}}(\gamma_{0}^{T}X_{i})^{T}\mathcal{C}_{0,k}X_i\epsilon_i+
\frac{1}{\sqrt{n}}\sum_{i=1}^{n}
(\widehat{\mathcal{C}}_{k}-\mathcal{C}_{0,k})^{T}
\mathcal{\dot{H}}(\gamma_{0}^{T}X_{i})X_i\epsilon_i\\
&+&\frac{1}{\sqrt{n}}\sum_{i=1}^{n}
\delta_{0,k}(\gamma_0^{T}X_{i})
\mathcal{\dot{H}}(\gamma_{0}^{T}X_{i})^{T}\mathcal{C}_{0,k}X_i\\
&+&\frac{1}{\sqrt{n}}\sum_{i=1}^{n}
\delta_{0,k}(\gamma_0^{T}X_{i})
\mathcal{\dot{H}}(\gamma_{0}^{T}X_{i})^{T}(\widehat{\mathcal{C}}_{k}-\mathcal{C}_{0,k})X_i\\
&+&\frac{1}{\sqrt{n}}\sum_{i=1}^{n}(\mathcal{C}_{0,k}-\widehat{\mathcal{C}}_{k})^{T}
\mathcal{H}(\gamma_0^{T}X_{i})
\mathcal{\dot{H}}(\gamma_{0}^{T}X_{i})^{T}(\widehat{\mathcal{C}}_{k}-\mathcal{C}_{0,k})X_i\\
&+&\frac{1}{\sqrt{n}}\sum_{i=1}^{n}(\mathcal{C}_{0,k}-\widehat{\mathcal{C}}_{k})^{T}
\mathcal{H}(\gamma_0^{T}X_{i})
\mathcal{\dot{H}}(\gamma_{0}^{T}X_{i})^{T}\mathcal{C}_{0,k}X_i\\
&=&\bar{J}_{n,11}+\bar{J}_{n,12}+\bar{J}_{n,13}+\bar{J}_{n,14}+\bar{J}_{n,15}+\bar{J}_{n,16}
\end{eqnarray*}
where the definitions of $\bar{J}_{n,11}$, $\bar{J}_{n,12}$, $\bar{J}_{n,13}$, $\bar{J}_{n,14}$,
$\bar{J}_{n,15}$ and $\bar{J}_{n,16}$
are obvious.

For $\bar{J}_{n,12}$, we have
\begin{eqnarray*}
E\left\|\bar{J}_{n,12}\right\|^{2}&=&E\left\|\frac{1}{\sqrt{n}}\sum_{i=1}^{n}
(\widehat{\mathcal{C}}_{k}-\mathcal{C}_{0,k})^{T}
\mathcal{\dot{H}}(\gamma_{0}^{T}X_{i})X_i\epsilon_i\right\|^{2}\\
&=&\frac{1}{n}\sum_{i=1}^{n}\sigma^{2}E\left\|(\widehat{\mathcal{C}}_{k}-\mathcal{C}_{0,k})^{T}
\mathcal{\dot{H}}(\gamma_{0}^{T}X_{i})X_i\right\|^{2}\\
&=&\frac{1}{n}\sum_{i=1}^{n}\sigma^{2}(\widehat{\mathcal{C}}_{k}-\mathcal{C}_{0,k})^{T}
E\left[\mathcal{\dot{H}}(\gamma_{0}^{T}X_{i})\mathcal{\dot{H}}(\gamma_{0}^{T}X_{i})^{T}\|XX^{T}\|^{2}\right]
(\widehat{\mathcal{C}}_{k}-\mathcal{C}_{0,k})\\
&\leq & O(1)\|\widehat{\mathcal{C}}_{k}-\mathcal{C}_{0,k}\|^{2}\leq O(1)\|\widehat{g}_k-g_{0}\|_{L^2}^{2}.
\end{eqnarray*}
Hence, we can obtain $\bar{J}_{n,12}=o_p(1)$.

For $\bar{J}_{n,13}$, we have
\begin{eqnarray*}
\left\|
\frac{1}{n}\sum_{i=1}^{n}
\delta_{0,k}(\gamma_0^{T}X_{i})
\mathcal{\dot{H}}(\gamma_{0}^{T}X_{i})^{T}\mathcal{C}_{0,k}X_i\right\|\leq \frac{1}{n}\sum_{i=1}^{n}\left\|
\delta_{0,k}(\gamma_0^{T}X_{i})
\mathcal{\dot{H}}(\gamma_{0}^{T}X_{i})^{T}\mathcal{C}_{0,k}X_i\right\|\\
\leq\left\{ \frac{1}{n}\sum_{i=1}^{n}\|\delta_{0,k}(\gamma_0^{T}X_{i})\|^{2} \right\}^{1/2}
\times \left\{ \frac{1}{n}\sum_{i=1}^{n}\|\mathcal{\dot{H}}(\gamma_{0}^{T}X_{i})^{T}\mathcal{C}_{0,k}X_i\|^{2} \right\}^{1/2}
=O_p(k^{-m/2}),
\end{eqnarray*}
where  the last equality follows from the facts that $E\|\delta_{0,k}(\gamma_0^{T}X)\|^{2}=O(k^{-m})$ and
$E\|\mathcal{\dot{H}}(\gamma_{0}^{T}X)^{T}\mathcal{C}_{0,k}X\|^{2}=O(1)$.
Hence, $\bar{J}_{n,13}=O_p(\sqrt{nk^{-m}})=o_p(1)$.
Using similar arguments, we can show that
$\bar{J}_{n,14}$ and  $\bar{J}_{n,15}$
are all $o_p(1)$.

For $\bar{J}_{n,2}$, we have
\begin{eqnarray*}
\bar{J}_{n,2}&=&	\frac{1}{\sqrt{n}}\sum_{i=1}^{n}
	\left(\widehat{Y}_{i}-Y_{i}^{*}\right)\mathcal{\dot{H}}
	(\gamma_{0}^{T}X_{i})^{T}\mathcal{C}_{0,k}X_i\\
&=&\frac{-1}{\sqrt{n}}\sum_{i=1}^{n}
\left[\frac{D_i(Y_i-\mu_{1}(X_i;\alpha_{1,0}))}{\pi^{2}(X_i;\beta^{*})}
\right]
\mathcal{\dot{H}}
(\gamma_{0}^{T}X_{i})^{T}\mathcal{C}_{0,k}X_i\pi^{(1)}(X_i;\beta^{*})^{T}(\widehat{\beta}-\beta_0)\\
&+&\frac{1}{\sqrt{n}}\sum_{i=1}^{n}\left[1-\frac{D_i}{\pi(X_i;\beta_0)}\right]
\mathcal{\dot{H}}
(\gamma_{0}^{T}X_{i})^{T}\mathcal{C}_{0,k}X_i\mu_{1}^{(1)}(X_i;\alpha_{1}^{*})^{T}(\widehat{\alpha}_1-\alpha_{1,0})\\
&+&\frac{1}{\sqrt{n}}\sum_{i=1}^{n}\frac{D_i(\widehat{\beta}-\beta_0)^{T}\pi^{1}(X_i;\beta^{*})}{\pi^{2}(X_i;\beta^{*})}
\mathcal{\dot{H}}
(\gamma_{0}^{T}X_{i})^{T}\mathcal{C}_{0,k}X_i\mu_{1}^{(1)}(X_i;\alpha_{1}^{*})^{T}(\widehat{\alpha}_1-\alpha_{1,0})\\
&+&\frac{-1}{\sqrt{n}}\sum_{i=1}^{n}
\left[\frac{(1-D_i)(Y_i-\mu_{0}(X_i;\alpha_{0,0}))}{(1-\pi(X_i;\beta^{*}))^{2}}
\right]
\mathcal{\dot{H}}
(\gamma_{0}^{T}X_{i})^{T}\mathcal{C}_{0,k}X_i\pi^{(1)}(X_i;\beta^{*})^{T}(\widehat{\beta}-\beta_0)\\
&+&\frac{1}{\sqrt{n}}\sum_{i=1}^{n}\left[\frac{1-D_i}{1-\pi(X_i;\beta_0)}-1\right]
\mathcal{\dot{H}}
(\gamma_{0}^{T}X_{i})^{T}\mathcal{C}_{0,k}X_i\mu_{0}^{(1)}(X_i;\alpha_{0}^{*})^{T}(\widehat{\alpha}_0-\alpha_{0,0})\\
&+&\frac{1}{\sqrt{n}}\sum_{i=1}^{n}\left[\frac{(1-D_i)(\widehat{\beta}-\beta_0)^{T}\pi^{(1)}(X_i;\beta^{*})}{(1-\pi(X_i;\beta^{*}))^{2}}\right]
\mathcal{\dot{H}}
(\gamma_{0}^{T}X_{i})^{T}\mathcal{C}_{0,k}X_i\mu_{0}^{(1)}(X_i;\alpha_{0}^{*})^{T}(\widehat{\alpha}_0-\alpha_{0,0})\\
&=&o_p(1),
\end{eqnarray*}
where
$\beta^{*}\in (\widehat{\beta}, \beta_0)$, $\alpha_{1}^{*}\in (\widehat{\alpha}_{1}, \alpha_{1,0})$ and  $\alpha_{0}^{*}\in (\widehat{\alpha}_{0}, \alpha_{0,0})$ are intermediate values.

For $\bar{J}_{n,3}$, we have
\begin{eqnarray*}
\bar{J}_{n,3}&=&\gamma_0\widehat{\gamma}^{T}\frac{1}{\sqrt{n}}\sum_{i=1}^{n}
\left[\mathcal{H}(\gamma_{0}^{T}X_{i})^{T}\mathcal{C}_{0,k}-\mathcal{H}(\widehat{\gamma}^{T}X_{i})^{T}\widehat{\mathcal{C}}_{k}\right]
\mathcal{\dot{H}}(\widehat{\gamma}^{T}X_{i})^{T}\widehat{\mathcal{C}}_{k}X_{i}\\
&+&\gamma_0\widehat{\gamma}^{T}\frac{1}{\sqrt{n}}\sum_{i=1}^{n}
\mathcal{\dot{H}}(\gamma_{0}^{T}X_{i})^{T}\mathcal{C}_{0,k}X_{i}\epsilon_i+
\gamma_0\widehat{\gamma}^{T}\frac{1}{\sqrt{n}}\sum_{i=1}^{n}
\delta_{0,k}(\gamma_0^{T}X_{i})
\mathcal{\dot{H}}(\widehat{\gamma}^{T}X_{i})^{T}\widehat{\mathcal{C}}_{k}X_{i}
\\
&+&\gamma_0\widehat{\gamma}^{T}\frac{1}{\sqrt{n}}\sum_{i=1}^{n}
[\mathcal{\dot{H}}(\widehat{\gamma}^{T}X_{i})^{T}\widehat{\mathcal{C}}_{k}-
\mathcal{\dot{H}}(\gamma_{0}^{T}X_{i})^{T}\mathcal{C}_{0,k}]X_{i}\epsilon_i
+\gamma_0\widehat{\gamma}^{T}\frac{1}{\sqrt{n}}\sum_{i=1}^{n}
[\widehat{Y}_{i}-Y_{i}^{*}]
\mathcal{\dot{H}}(\widehat{\gamma}^{T}X_{i})^{T}\widehat{\mathcal{C}}_{k}X_{i}\\
&=&\gamma_0\widehat{\gamma}^{T}\bar{J}_{n,31}+\gamma_0\widehat{\gamma}^{T}\bar{J}_{n,32}+\gamma_0\widehat{\gamma}^{T}\bar{J}_{n,33}+
\gamma_0\widehat{\gamma}^{T}\bar{J}_{n,34}+\gamma_0\widehat{\gamma}^{T}\bar{J}_{n,35},
\end{eqnarray*}
where the definitions of $\bar{J}_{n,31}$, $\bar{J}_{n,32}$, $\bar{J}_{n,33}$, $\bar{J}_{n,34}$, and $\bar{J}_{n,35}$
are obvious.

For $\bar{J}_{n,31}$, we have
\begin{eqnarray*}
\bar{J}_{n,31}&=&\frac{1}{\sqrt{n}}\sum_{i=1}^{n}
\left[\mathcal{H}(\gamma_{0}^{T}X_{i})^{T}\mathcal{C}_{0,k}-\mathcal{H}(\gamma_{0}^{T}X_{i})^{T}\widehat{\mathcal{C}}_{k}\right]
\mathcal{\dot{H}}(\widehat{\gamma}^{T}X_{i})^{T}\widehat{\mathcal{C}}_{k}X_{i}\\
&+&\frac{1}{\sqrt{n}}\sum_{i=1}^{n}
\left[\mathcal{H}(\gamma_{0}^{T}X_{i})^{T}\widehat{\mathcal{C}}_{k}-\mathcal{H}(\widehat{\gamma}^{T}X_{i})^{T}\widehat{\mathcal{C}}_{k}\right]
\mathcal{\dot{H}}(\widehat{\gamma}^{T}X_{i})^{T}\widehat{\mathcal{C}}_{k}X_{i}\\
&=&\sqrt{n}(\mathcal{C}_{0,k}-\widehat{\mathcal{C}}_{k})^{T}E[
\mathcal{\dot{H}}(\gamma_{0}^{T}X)^{T}\mathcal{\dot{H}}(\gamma_{0}^{T}X)^{T}\mathcal{C}_{0,k}X]\\
&+&E[
(\mathcal{\dot{H}}(\gamma_{0}^{T}X)^{T}\mathcal{C}_{0,k})^{2}XX^{T}]\sqrt{n}(\gamma_0-\widehat{\gamma})+o_p(1).
\end{eqnarray*}

For $\bar{J}_{n,33}$, we have
\begin{eqnarray*}
\bar{J}_{n,33}&=&\frac{1}{\sqrt{n}}\sum_{i=1}^{n}
\delta_{0,k}(\gamma_0^{T}X_{i})
\mathcal{\dot{H}}(\gamma_{0}^{T}X_{i})^{T}\mathcal{C}_{0,k}X_{i}\\
&+&\frac{1}{\sqrt{n}}\sum_{i=1}^{n}\delta_{0,k}(\gamma_0^{T}X_{i})
(\widehat{\mathcal{C}}_{k}-\mathcal{C}_{0,k})^{T}
\mathcal{\dot{H}}(\widehat{\gamma}^{T}X_{i})X_{i}\\
&+&\frac{1}{\sqrt{n}}\sum_{i=1}^{n}\delta_{0,k}(\gamma_0^{T}X_{i})
(\widehat{\gamma}-\gamma_{0})^{T}X_i
\mathcal{\ddot{H}}(\gamma_{*}^{T}X_{i})^{T}\mathcal{C}_{0,k}
X_{i}\\
&=&\bar{J}_{n,331}+\bar{J}_{n,332}+\bar{J}_{n,333},
\end{eqnarray*}
where the definitions of $\bar{J}_{n,331}$, $\bar{J}_{n,332}$ and $\bar{J}_{n,333}$  are obvious.
For $\bar{J}_{n,331}$, we have the  following expression
\begin{eqnarray*}
E\left\|\bar{J}_{n,331}\right\|^{2}&=&E\left\|\frac{1}{\sqrt{n}}\sum_{i=1}^{n}
\delta_{0,k}(\gamma_0^{T}X_{i})
\mathcal{\dot{H}}(\gamma_{0}^{T}X_{i})^{T}\mathcal{C}_{0,k}X_{i}\right\|^{2}\\
&\leq & O(1) E\left\| \delta_{0,k}(\gamma_0^{T}X)\mathcal{\dot{H}}(\gamma_{0}^{T}X)^{T}\mathcal{C}_{0,k}X\right\|^{2}\\
&\leq & O(1)\{E|\delta_{0,k}(\gamma_0^{T}X)|^{4}E\|\mathcal{\dot{H}}(\gamma_{0}^{T}X)^{T}\mathcal{C}_{0,k}X\|^{4}\}^{1/2}\\
&=&O(1)(E|\delta_{0,k}(\gamma_0^{T}X)|^{4})^{1/2}=o(1).
\end{eqnarray*}
Hence, $\bar{J}_{n,331}=o_p(1)$. Using similar arguments, we can show that
$\bar{J}_{n,332}$ and  $\bar{J}_{n,333}$
are both $o_p(1)$.

For $\bar{J}_{n,34}$, we have
\begin{eqnarray*}
\bar{J}_{n,34}&=&\frac{1}{\sqrt{n}}\sum_{i=1}^{n}
[\mathcal{\dot{H}}(\widehat{\gamma}^{T}X_{i})^{T}\widehat{\mathcal{C}}_{k}-
\mathcal{\dot{H}}(\widehat{\gamma}^{T}X_{i})^{T}C_{0,k}]X_{i}\epsilon_i\\
&+&\frac{1}{\sqrt{n}}\sum_{i=1}^{n}
[\mathcal{\dot{H}}(\widehat{\gamma}^{T}X_{i})^{T}\mathcal{C}_{0,k}-
\mathcal{\dot{H}}(\gamma_{0}^{T}X_{i})^{T}\mathcal{C}_{0,k}]X_{i}\epsilon_i\\
&=&\bar{J}_{n,341}+\bar{J}_{n,342},
\end{eqnarray*}
where the definitions of $\bar{J}_{n,341}$ and $\bar{J}_{n,342}$  are obvious.
For term $\bar{J}_{n,341}$, we have
\begin{eqnarray*}
E\left\| \frac{1}{\sqrt{n}}\sum_{i=1}^{n}
(\widehat{\mathcal{C}}_{k}-\mathcal{C}_{0,k})^{T}
\mathcal{\dot{H}}(\widehat{\gamma}^{T}X_{i})X_{i}\epsilon_i\right\|^{2}
=\frac{1}{n}\sum_{i=1}^{n}\sigma^{2}E\left\|\widehat{\mathcal{C}}_{k}-\mathcal{C}_{0,k})^{T}
\mathcal{\dot{H}}(\widehat{\gamma}^{T}X_{i})X_{i}\right\|^{2}\\
\leq O(1)\|\widehat{\mathcal{C}}_{k}-\mathcal{C}_{0,k}\|^{2}=o_p(1).
\end{eqnarray*}
Thus,  $\bar{J}_{n,341}=o_p(1)$, and similarly  $\bar{J}_{n,342}=o_p(1)$. Furthermore,
$\bar{J}_{n,34}=o_p(1)$.

For $\bar{J}_{n,35}$, we have
\begin{eqnarray*}
\bar{J}_{n,35}&=&\frac{1}{\sqrt{n}}\sum_{i=1}^{n}
(\widehat{Y}_{i}-Y_{i}^{*})
\mathcal{\dot{H}}(\gamma_{0}^{T}X_{i})^{T}\mathcal{C}_{0,k}X_{i}\\
&+&\frac{1}{\sqrt{n}}\sum_{i=1}^{n}
(\widehat{Y}_{i}-Y_{i}^{*})(\widehat{\gamma}-\gamma_0)^{T}X_i
\mathcal{\ddot{H}}(\gamma_{*}^{T}X_{i})^{T}
\widehat{\mathcal{C}}_{k}X_{i}\\
&+&\frac{1}{\sqrt{n}}\sum_{i=1}^{n}
(\widehat{Y}_{i}-Y_{i}^{*})
\mathcal{\dot{H}}(\gamma_{0}^{T}X_{i})^{T}(\widehat{\mathcal{C}}_{k}-\mathcal{C}_{0,k})
X_{i}
=o_p(1),
\end{eqnarray*}
where the convergence result follows from arguments similar to those used for analyzing $\bar{J}_{n,2}$.

Hence, $\sqrt{n} U_{n1}(\gamma_0,\widehat{\mathcal{C}}_{k},\widehat{\lambda})$ can be re-written as:
\begin{eqnarray*}
\sqrt{n} U_{n1}(\gamma_0,\widehat{\mathcal{C}}_{k},\widehat{\lambda})&=&
(-I_{p}+\gamma_{0}\widehat{\gamma}^{T})\frac{2}{\sqrt{n}}\sum_{i=1}^{n}
\mathcal{\dot{H}}(\gamma_{0}^{T}X_{i})^{T}\mathcal{C}_{0,k}X_i\epsilon_i\\
&-&\gamma_{0}\widehat{\gamma}^{T}2E\left[
(\mathcal{\dot{H}}(\gamma_{0}^{T}X)^{T}\mathcal{C}_{0,k})^{2}XX^{T}\right]\sqrt{n}(\widehat{\gamma}-\gamma_0)+o_p(1).
\end{eqnarray*}

Due to the fact that
\begin{eqnarray*}
\widehat{\gamma}^{T}\gamma_0-1&=& (\widehat{\gamma}-\gamma_0)^{T}\gamma_0=(\widehat{\gamma}-\gamma_0)^{T}(\gamma_0-\widehat{\gamma}+\widehat{\gamma})\\
&=&-\|\widehat{\gamma}-\gamma_0\|^{2}+1-\widehat{\gamma}^{T}\gamma_0,
\end{eqnarray*}
we have  the identity $1-\widehat{\gamma}^{T}\gamma_0=\frac{1}{2}\|\widehat{\gamma}-\gamma_0\|^{2}$,
where  the left-hand side
 $1-\widehat{\gamma}^{T}\gamma_0$  quantifies the deviation of the estimated parameter $\widehat{\gamma}$
 from the true parameter $\gamma_0$ by measuring its distance to the surface of the unit sphere.
 This equality reveals that the convergence rate of $\widehat{\gamma}$  is accelerated along the direction of $\gamma_0$
 relative to all orthogonal directions.

By using Taylor expansion, we can obtain
\begin{eqnarray*}
0&=&V^{T}\sqrt{n}U_{n1}(\widehat{\gamma},\widehat{\mathcal{C}}_{k},\widehat{\lambda})\\
&=&V^{T}\sqrt{n}U_{n1}(\gamma_0,\widehat{\mathcal{C}}_{k},\widehat{\lambda})+V^{T}\frac{\partial U_{n,1}(\bar{\gamma},\widehat{\mathcal{C}}_{k},\widehat{\lambda})}
{\partial\gamma}\sqrt{n}(\widehat{\gamma}-\gamma_0)\\
&=&V^{T}\sqrt{n}U_{n1}(\gamma_0,\widehat{\mathcal{C}}_{k},\widehat{\lambda})+V^{T}\frac{\partial U_{n,1}(\bar{\gamma},\widehat{\mathcal{C}}_{k},\widehat{\lambda})}
{\partial\gamma}\sqrt{n}(VV^{T}+\gamma_{0}\gamma_{0}^{T})(\widehat{\gamma}-\gamma_0)\\
&=&V^{T}\sqrt{n}U_{n1}(\gamma_0,\widehat{\mathcal{C}}_{k},\widehat{\lambda})+V^{T}\frac{\partial U_{n,1}(\bar{\gamma},\widehat{\mathcal{C}}_{k},\widehat{\lambda})}
{\partial\gamma}V\sqrt{n}V^{T}(\widehat{\gamma}-\gamma_0)\\
&-&\frac{1}{2\sqrt{n}}\|\sqrt{n}(\widehat{\gamma}-\gamma_0)\|^{2}
V^{T}\frac{\partial U_{n,1}(\bar{\gamma},\widehat{\mathcal{C}}_{k},\widehat{\lambda})}
{\partial\gamma}\gamma_0.
\end{eqnarray*}
By the Central Limit Theorem, it follows that
\begin{eqnarray*}
\sqrt{n}V^{T}(\widehat{\gamma}-\gamma_0)\xrightarrow{d}\mathcal{N}(0,\sigma^2(V^{T}\Sigma_{\Gamma} V)^{-1})
\end{eqnarray*}
with $\Sigma_{\Gamma}=E\left[\mathcal{C}_{0,k}^{T}\mathcal{\dot{H}}(\gamma_{0}^{T}X_{i})
\mathcal{\dot{H}}(\gamma_{0}^{T}X_{i})^{T}\mathcal{C}_{0,k}XX^{T}\right]$. Hence, Theorem 2 holds.

{\bf Proof of Theorem 3}\quad
We will focus on the term $\frac{\partial U_{n,2}(\gamma,\mathcal{C}_{k}) }
{\partial \gamma}|_{(\gamma,\mathcal{C}_{k})=(\widehat{\gamma},\bar{\mathcal{C}}_{k})}$
with $\bar{\mathcal{C}}_{k}$ lies between $\widehat{\mathcal{C}}_{k}$ and $\mathcal{C}_{0,k}$, where
\begin{eqnarray*}
\frac{\partial U_{n,2}(\gamma,\mathcal{C}_{k}) }{\partial \mathcal{C}_{k}}	&=&\frac{2}{n}\sum_{i=1}^{n}
\mathcal{\dot{H}}(\gamma^{T}X_{i})
\mathcal{\dot{H}}(\gamma^{T}X_{i})^{T}.
\end{eqnarray*}
By the law of large numbers and continuous mapping theory, we have
\begin{eqnarray*}
\frac{\partial U_{n,2}(\widehat{\gamma},\mathcal{C}_{k}) }{\partial \mathcal{C}_{k}}=2E\left[\mathcal{\dot{H}}(\gamma_{0}^{T}X)
\mathcal{\dot{H}}(\gamma_{0}^{T}X)^{T}\right]+o_p(1).
\end{eqnarray*}
Expand $\sqrt{n} U_{n2}(\gamma,\mathcal{C}_{k})|_{(\gamma,\mathcal{C}_{k})
=(\widehat{\gamma},\mathcal{C}_{0,k})}$ as follows:
\begin{eqnarray*}
	\sqrt{n} U_{n,2}(\widehat{\gamma},\mathcal{C}_{0,k})&=&\frac{-2}{\sqrt{n}}\sum_{i=1}^{n}	\left[\mathcal{H}(\gamma_{0}^{T}X_{i})^{T}\mathcal{C}_{0,k}-\mathcal{H}(\widehat{\gamma}^{T}X_{i})^{T}\mathcal{C}_{0,k}\right]
\mathcal{H}(\widehat{\gamma}^{T}X_{i})\\
&+&\frac{-2}{\sqrt{n}}\sum_{i=1}^{n}	\delta_{0,k}(\gamma_{0}^{T}X_{i})\mathcal{H}(\widehat{\gamma}^{T}X_{i})
+\frac{-2}{\sqrt{n}}\sum_{i=1}^{n}\mathcal{H}(\widehat{\gamma}^{T}X_{i})\epsilon_i
\\
&+&\frac{-2}{\sqrt{n}}\sum_{i=1}^{n}
	[\widehat{Y}_{i}-Y_{i}^{*}]\mathcal{H}(\widehat{\gamma}^{T}X_{i})\\
&=&-2\widetilde{J}_{n,1}-2\widetilde{J}_{n,2}-2\widetilde{J}_{n,3}-2\widetilde{J}_{n,4},
\end{eqnarray*}
where the definitions of $\widetilde{J}_{n,1}$,  $\widetilde{J}_{n,2}$,  $\widetilde{J}_{n,3}$ and $\widetilde{J}_{n,4}$  are obvious.

For $\widetilde{J}_{n,1}$, we have
\begin{eqnarray*}
\widetilde{J}_{n,1}&=&\frac{1}{\sqrt{n}}\sum_{i=1}^{n}	\left[\mathcal{H}(\gamma_{0}^{T}X_{i})^{T}\mathcal{C}_{0,k}-\mathcal{H}(\widehat{\gamma}^{T}X_{i})^{T}\mathcal{C}_{0,k}\right]\times
\left[\mathcal{H}(\widehat{\gamma}^{T}X_{i})-\mathcal{H}(\gamma_{0}^{T}X_{i})\right]\\
&+&\frac{1}{\sqrt{n}}\sum_{i=1}^{n}	\left[\mathcal{H}(\gamma_{0}^{T}X_{i})^{T}\mathcal{C}_{0,k}-\mathcal{H}(\widehat{\gamma}^{T}X_{i})^{T}\mathcal{C}_{0,k}\right]
\mathcal{H}(\gamma_{0}^{T}X_{i})\\
&=&\frac{1}{\sqrt{n}}\sum_{i=1}^{n}	(\gamma_0-\widehat{\gamma})^{T}X_i\mathcal{\dot{H}}(\bar{\gamma}^{T}X_{i})^{T}\mathcal{C}_{0,k}
\mathcal{\dot{H}}(\bar{\gamma}^{T}X_{i})X_{i}^{T}(\gamma_0-\widehat{\gamma})\\
&+&\frac{1}{n}\sum_{i=1}^{n}\mathcal{\dot{H}}(\bar{\gamma}^{T}X_{i})^{T}\mathcal{C}_{0,k}
\mathcal{H}(\gamma_{0}^{T}X_{i})X_{i}^{T}\sqrt{n}(\gamma_0-\widehat{\gamma})\\
&=&E\left[\mathcal{\dot{H}}(\gamma_0^{T}X)^{T}\mathcal{C}_{0,k}
\mathcal{H}(\gamma_{0}^{T}X)X^{T}\right]\sqrt{n}(\gamma_0-\widehat{\gamma})+o_p(1).
\end{eqnarray*}

Following arguments similar to those used for $\bar{J}_{n,33}$ and $\bar{J}_{n,35}$,
 we  obtain
$\widetilde{J}_{n,2}=o_p(1)$ and $\widetilde{J}_{n,4}=o_p(1)$.
For $\widetilde{J}_{n,3}$, we have
\begin{eqnarray*}
\widetilde{J}_{n,3}&=&\frac{1}{\sqrt{n}}\sum_{i=1}^{n}\mathcal{H}(\gamma_{0}^{T}X_{i})\epsilon_i
+\frac{1}{n}\sum_{i=1}^{n}\epsilon_i\mathcal{\dot{H}}(\bar{\gamma}^{T}X_{i})X_i^{T}\sqrt{n}(\widehat{\gamma}-\gamma_0)\\
&=&\frac{1}{\sqrt{n}}\sum_{i=1}^{n}\mathcal{H}(\gamma_{0}^{T}X_{i})\epsilon_i+o_p(1).
\end{eqnarray*}

By using Taylor expansion, we  obtain
\begin{eqnarray*}
0&=&\sqrt{n}U_{n2}(\widehat{\gamma},\widehat{\mathcal{C}}_{k})\\
&=&\sqrt{n}U_{n2}(\widehat{\gamma},\mathcal{C}_{0,k})+\frac{\partial U_{n,2}(\widehat{\gamma},\bar{\mathcal{C}}_{k}) }{\partial \mathcal{C}_{k}}
\sqrt{n}(\widehat{\mathcal{C}}_{k}-\mathcal{C}_{0,k})\\
&=&\frac{\partial U_{n,2}(\widehat{\gamma},\bar{\mathcal{C}}_{k}) }{\partial \mathcal{C}_{k}}
\sqrt{n}(\widehat{\mathcal{C}}_{k}-\mathcal{C}_{0,k})-2E\left[\mathcal{\dot{H}}(\gamma_0^{T}X)^{T}\mathcal{C}_{0,k}
\mathcal{H}(\gamma_{0}^{T}X)X^{T}\right]V\sqrt{n}V^{T}(\gamma_0-\widehat{\gamma})\\
&+&\frac{1}{\sqrt{n}}\|\sqrt{n}(\widehat{\gamma}-\gamma_0)\|^{2}E\left[\mathcal{\dot{H}}(\gamma_0^{T}X)^{T}\mathcal{C}_{0,k}
\mathcal{H}(\gamma_{0}^{T}X)X^{T}\right]\gamma_0\\
&-&\frac{2}{\sqrt{n}}\sum_{i=1}^{n}\mathcal{H}(\gamma_{0}^{T}X_{i})\epsilon_i+o_p(1).
\end{eqnarray*}
 By the Central Limit Theorem, it follows that
\begin{eqnarray*}
\sqrt{n}(\widehat{\mathcal{C}}_{k}-\mathcal{C}_{0,k})\xrightarrow{d}\mathcal{N}(0,\sigma^2\Sigma_{A}^{-1}\Sigma_{\mathcal{C}}\Sigma_{A}^{-1}),
\end{eqnarray*}
where
\begin{eqnarray*}
\Sigma_{A}&=&E\left[\mathcal{\dot{H}}(\gamma_{0}^{T}X)
\mathcal{\dot{H}}(\gamma_{0}^{T}X)^{T}\right],
\Sigma_{B}=E\left[\mathcal{H}(\gamma_{0}^{T}X)\mathcal{H}(\gamma_{0}^{T}X)^{T}\right],\\
\Sigma_{D}&=&E\left[\mathcal{\dot{H}}(\gamma_0^{T}X)^{T}\mathcal{C}_{0,k}
\mathcal{H}(\gamma_{0}^{T}X)X^{T}\right], \Sigma_{\mathcal{C}}=\Sigma_{B}+
\Sigma_{D}V(V^{T}\Sigma_{\Gamma}V)^{-1}V^{T}\Sigma_{D}^{T}.
\end{eqnarray*}

\newpage
\begin{center}
\begin{table}
\caption{Simulation results }
\vspace{1mm}
\begin{center}
\begin{tabular}{ccrccccrccc}
\hline \hline

&& \multicolumn{4}{c}{Prop=0.3} & \multicolumn{4}{c}{Prop=0.5}
\\
\cline{3-6} \cline{8-11}
 $n$ & Estimator     & Bias   &  SD  &  ESE  & CI  & & Bias   &  SD  &  ESE    & CI \\
\hline
&&&&& $g_{0}(x)=x$\\
1000 & $\widehat{\gamma}_1$ & -0.004 & 0.053  &0.059 &0.960  && -0.005  &0.052  &0.053  &0.946\\
&        $\widehat{\gamma}_2$ & 0.008 & 0.070 &0.077   &0.964  && 0.007 &0.075   &0.070  &0.947\\
&        $\widehat{\gamma}_3$ & 0.003 & 0.091 &0.093  &0.947  && 0.001 &0.086   &0.083  &0.941\\

1500 &  $\widehat{\gamma}_1$ & -0.005  & 0.044  &0.070  &0.953  && -0.004  &0.042  &0.043  &0.959\\
&        $\widehat{\gamma}_2$ & 0.002 & 0.059 &0.061   &0.958  && 0.003 &0.056   &0.057  &0.955\\
&        $\widehat{\gamma}_3$ & 0.002 & 0.073 &0.075   &0.957  && 0.001 &0.070   &0.068  &0.945\\
&&&&& $g_{0}(x)=2x^3-1$\\
1000 & $\widehat{\gamma}_1$ & -0.005 & 0.040  &0.037 &0.951  && -0.016  &0.037  &0.034  &0.955\\
&        $\widehat{\gamma}_2$ & -0.001 & 0.048 &0.047   &0.945  && -0.015 &0.048   &0.043  &0.939\\
&        $\widehat{\gamma}_3$ & 0.003 & 0.061 &0.058  &0.959  && 0.018 &0.060   &0.054  &0.945\\

1500 &  $\widehat{\gamma}_1$ & -0.003  & 0.029  &0.029  &0.943  && -0.013  &0.031  &0.029  &0.944\\
&        $\widehat{\gamma}_2$ & -0.001 & 0.037 &0.037   &0.939  && -0.013 &0.041   &0.037  &0.930\\
&        $\widehat{\gamma}_3$ & 0.003 & 0.047 &0.047   &0.956  && 0.017 &0.052   &0.047  &0.942\\
 \hline\hline

\end{tabular}
\end{center}
\vspace{2mm}
Notation: Prop: denotes the  proportion of the treatment assignment.
\end{table}
\end{center}

\begin{center}
\begin{figure}
  \centering
  \includegraphics[width=6.5in]{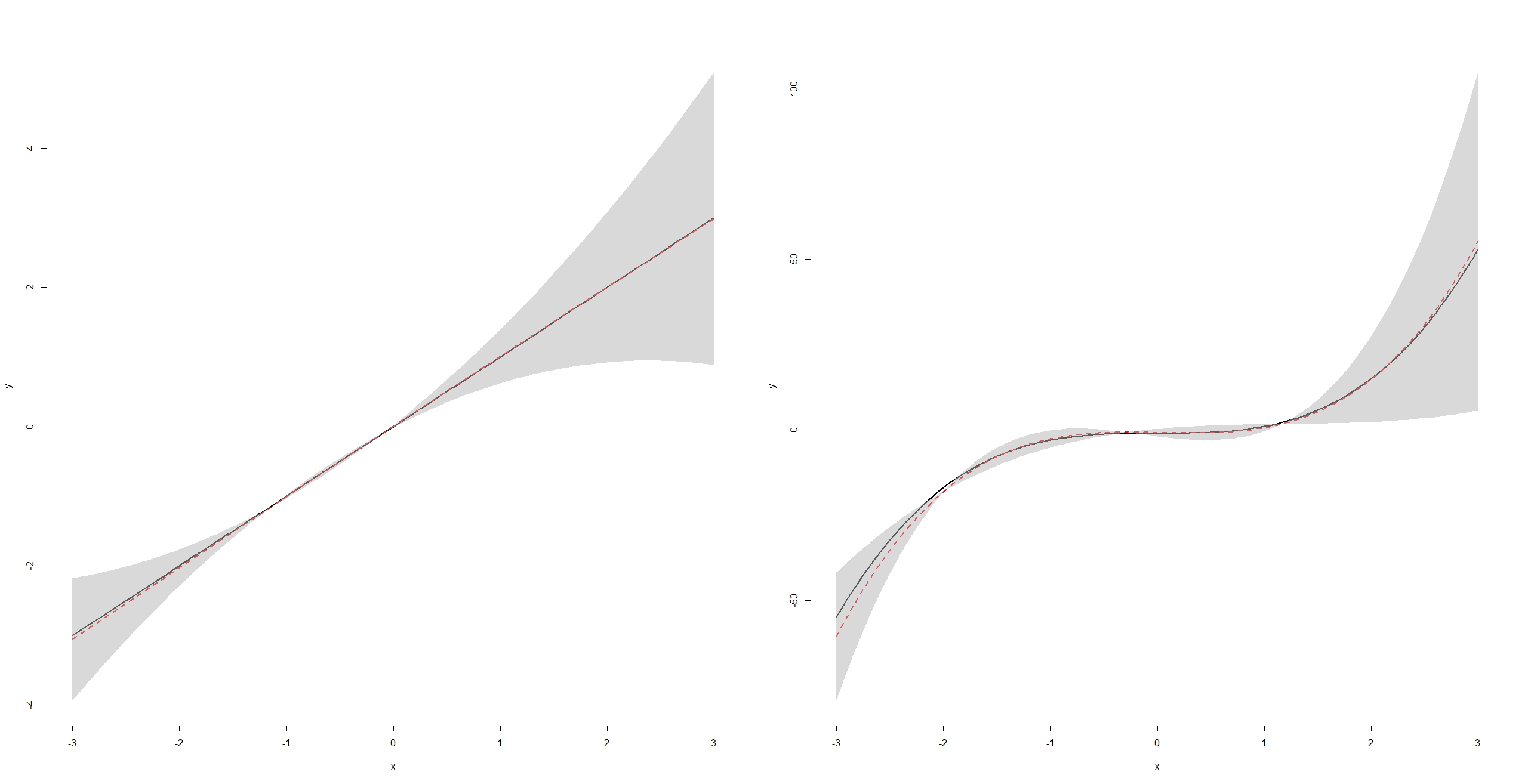}
 \caption{Estimates of the link function}
\end{figure}
\end{center}

\begin{center}
\begin{figure}
  \centering
  \includegraphics[width=4in]{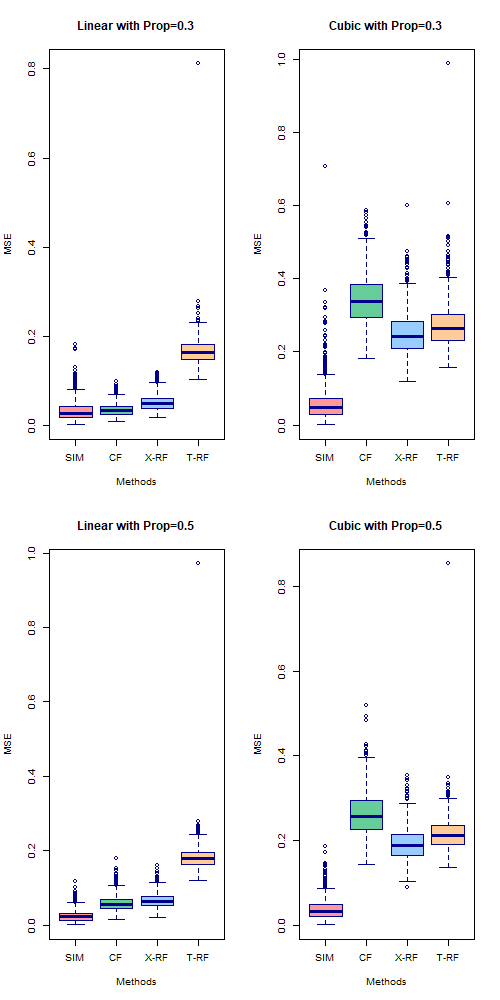}
 \caption{The mean squared errors of the four methods}
\end{figure}
\end{center}

\begin{center}
\begin{figure}
  \centering
  \includegraphics[width=4in]{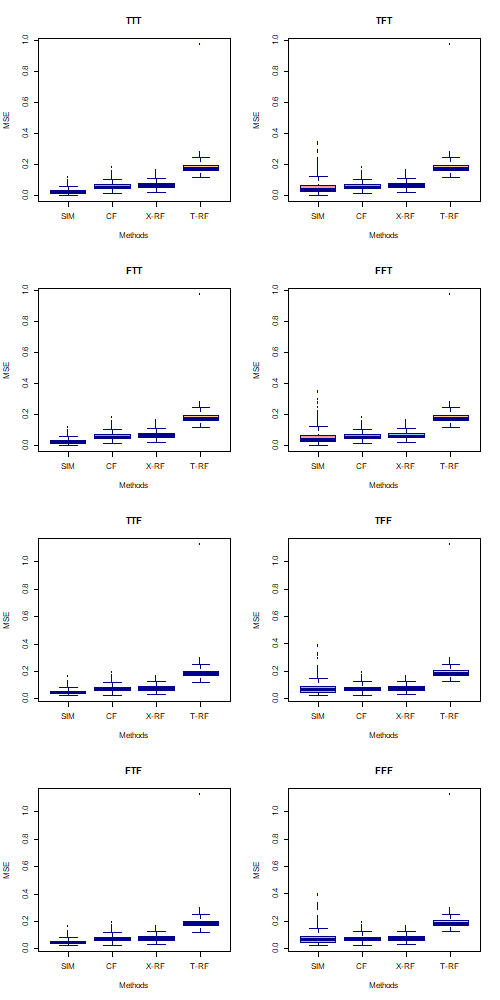}
 \caption{The mean squared errors of the four methods}
\end{figure}
\end{center}

\begin{center}
\begin{figure}
  \centering
  \includegraphics[width=4in]{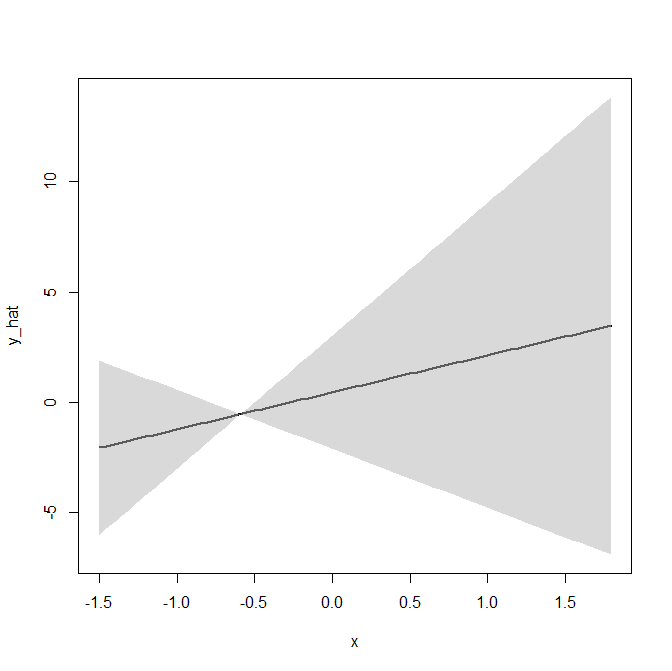}
 \caption{Estimated link function of NHANES}
\end{figure}
\end{center}

\begin{center}
\begin{figure}
  \centering
  \includegraphics[width=4in]{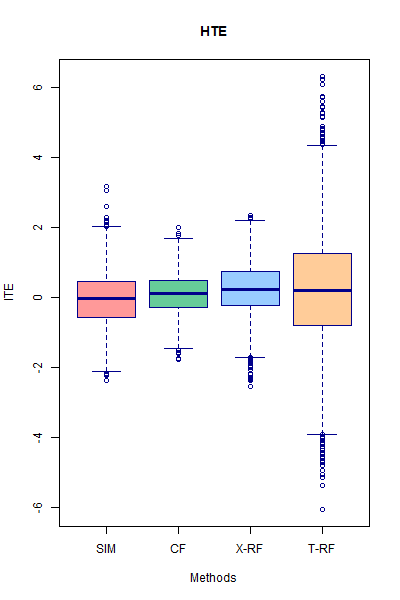}
 \caption{The individual treatment effects of the four methods}
\end{figure}
\end{center}

\begin{center}
\begin{table}
\caption{National Health and Nutrition Examination Survey }
\vspace{1mm}
\begin{center}
\begin{tabular}{crcc}
\hline \hline

Variable & Estimator &    SD  &  $p_{value}$  \\
\hline
age & 0.001 & 0.133 & 0.496\\
childsex & 0.011 & 0.254& 0.482\\
afam & -0.995  &0.356  &0.003\\
hisam & -0.009 &0.231  &0.485\\
$povlev_{200} $ & -0.008 &0.318 &0.490\\
$supnut_{prog}$ &-0.017 &0.417 & 0.484\\
$foodstamp_{prog}$ & -0.058 &0.391 &0.441\\
$foodsec_{chd}$ &-0.051  &0.290  &0.430\\
anyins &-0.057 &0.354 &0.436\\
refsex &0.008 &0.192 &0.483\\
refage &0.001 &0.100 &0.495\\

 \hline\hline

\end{tabular}
\end{center}
\end{table}
\end{center}


\newpage
\bibliographystyle{apalike}
\bibliography{HTE_Ref}

\end{document}